\documentclass[journal]{IEEEtran}

\usepackage[cmex10,intlimits]{amsmath}
\usepackage{amsfonts}
\usepackage{amssymb}
\usepackage{graphicx}
\usepackage{booktabs}
\usepackage{physics}

\usepackage{capekCommands}
\usepackage{capekColors}

\providecommand{\xv}{\ensuremath{\M{x}_\T{L}}}
\providecommand{\Gm}{\ensuremath{\V{\Gamma}}}

\providecommand{\Rmmat}{\ensuremath{\M{R}_\rho}}
\providecommand{\Rmvac}{\ensuremath{\M{R}_0}}
\providecommand{\Xmvac}{\ensuremath{\M{X}_0}}

\begin{document}

\title{A Role of Symmetries in Evaluation of Fundamental Bounds}
\author{Miloslav~Capek, \IEEEmembership{Senior Member, IEEE}, Lukas~Jelinek, and Michal~Masek
\thanks{Manuscript received \today; revised \today. This work was supported by the Czech Science Foundation under project~\mbox{No.~19-06049S} and by Ministry of Education, Youth and Sports under project~\mbox{LTAIN19047}.}
\thanks{M. Capek, L. Jelinek, and M. Masek are with the Czech Technical University in Prague, Prague, Czech Republic (e-mails: \{miloslav.capek; lukas.jelinek; michal.masek\}@fel.cvut.cz).}
\thanks{Color versions of one or more of the figures in this paper are
available online at http://ieeexplore.ieee.org.}
\thanks{Digital Object Identifier XXX}
}

\maketitle

\begin{abstract}
A problem of the erroneous duality gap caused by the presence of symmetries is solved in this paper utilizing point group theory. The optimization problems are first divided into two classes based on their predisposition to suffer from this deficiency. Then, the classical problem of Q-factor minimization is shown in an example where the erroneous duality gap is eliminated by combining solutions from orthogonal sub-spaces. Validity of this treatment is demonstrated in a series of subsequent examples of increasing complexity spanning the wide variety of optimization problems, namely minimum Q-factor, maximum antenna gain, minimum total active reflection coefficient, or maximum radiation efficiency with self-resonant constraint. They involve problems with algebraic and geometric multiplicities of the eigenmodes, and are completed by an example introducing the selective modification of modal currents falling into one of the symmetry-conformal sub-spaces. The entire treatment is accompanied with a discussion of finite numerical precision, and mesh grid imperfections and their influence on the results. Finally, the robust and unified algorithm is proposed and discussed, including advanced topics such as the uniqueness of the optimal solutions, dependence on the number of constraints, or an interpretation of the qualitative difference between the two classes of the optimization problems.
\end{abstract}


\section{Introduction}
\label{sec:Intro}

\IEEEPARstart{F}{undamental} bounds expressed in terms of source quantities~\cite{GustafssonTayliEhrenborgEtAl_AntennaCurrentOptimizationUsingMatlabAndCVX, JelinekCapek_OnTheStoredAndRadiatedEnergyDensity} have shown their versatility and usefulness for a wide range of applications in antenna theory, microwaves, and optics. They delimit the performance of theoretically feasible structures which help to judge the performance of existent designs~\cite{BestHanna_AperformanceComparisonOfFundamentalESA} and, in a few cases, lead to the conclusion that existing designs have already reached the bounds~\cite{Capek_etal_2019_OptimalPlanarElectricDipoleAntennas}. Additionally, given that the bounds are far from the actual performance of the devices became the driving force to search for better designs~\cite{Jelinek+etal2018}. However, despite recent success and a straightforward implementation, the problem with the presence of geometry symmetries remained open~\cite{GustafssonTayliEhrenborgEtAl_AntennaCurrentOptimizationUsingMatlabAndCVX, CapekGustafssonSchab_MinimizationOfAntennaQualityFactor, GustafssonCapek_MaximumGainEffAreaAndDirectivity}.

Under certain conditions, discussed in detail in this paper, a large class of optimization problems experience difficulties when symmetries are present. Although the problem is of a technical nature, it has a serious impact on the validity of the results since the degeneracy of eigenvalues introduces a duality gap between dual and primal solutions~\cite{BoydVandenberghe_ConvexOptimization}. This duality gap is manifested by the fact that the current solution for a primal was not constructed correctly. The known empirical solutions to this issue utilize an \textit{ad hoc} combination of the degenerated eigenvectors~\cite{CapekGustafssonSchab_MinimizationOfAntennaQualityFactor, GustafssonCapek_MaximumGainEffAreaAndDirectivity}. This approach is difficult to apply inside a general solver dealing with a large class of problems and structures of arbitrary geometry. The main difficulty, however, arises with structures of higher-order geometry degeneracies where the choice of modes to be combined is non-trivial. Since the shapes exhibiting symmetries are often used as initial designs, and since it is expected that the field of fundamental bounds will expand into a plethora of yet unsolved problems and researchers may face the problem again, a comprehensive and general treatment of this issue is of considerable importance.

The proposed solution adheres to point group theory, namely, the von Neuman-Wigner theorem~\cite{vonNeumannWigner_OnTheBehaviourOfEigenvaluesENG} is applied to a spectrum of eigenvalue traces given by the stationary points of the optimization problem. Consequently, the conditions under which the problem arises are discussed including how the problem is always connected to an underlying (parameterized) eigenvalue problem introducing an erroneous duality gap. A simple procedure showing how to detect when the problem occurs and how to close the erroneous duality gap is given. The proposed recipe can also treat cases of realistic mesh grids, \ie{}, those not perfectly respecting the symmetry groups of the original object. The procedure was thoroughly tested on many canonical objects, such as a rectangular plate, square plate, metallic rim with ground plane, in-parallel placed and crossed dipoles, spherical shell, etc.

The paper is organized as follows. The situation is thoroughly analyzed in Section~\ref{sec:QCQP}. It is realized that the erroneous duality gap occurs only when the eigenvalue solution is required, \ie{}, for quadratically constrained quadratic programs (QCQP) without linear terms. When the linear terms are present, this ambiguity vanishes as the solution does not use eigenvalue decomposition. The erroneous duality gap is closed in Section~\ref{sec:Symmetries} with the help of point group theory. In Section~\ref{sec:examples}, some examples are explicitly treated, showing where and how the symmetries appeared and what is their influence on the problem. The properties of the method are discussed in Section~\ref{sec:disc}. The uniqueness of the results (current density, port voltages, etc.) is investigated in light of the knowledge gained from the symmetry treatment. It is shown, that the presence of symmetries may introduce additional degrees of freedom for the optimization and it is specified where it is so. The paper is concluded in Section~\ref{sec:concl}.

\section{QCQP Problems}
\label{sec:QCQP}

The evaluation of the source quantity-based fundamental bounds starts with a statement of the optimization problem. Two problems, denoted as~$\OP{P}_1$ and~$\OP{P}_2$, are shown below to distinguish when the problem with symmetries may ($\OP{P}_1$) or may not ($\OP{P}_2$) arise. After establishing the Lagrangians, the optimization problems are solved via dual formulation~\cite{BoydVandenberghe_ConvexOptimization} the solution to which are subsequently interpreted with respect to point group theory.

\subsection{Optimization Problem~$\OP{P}_1$}
\label{subsec:QCQP1}

\begin{figure}
\centering
\includegraphics[width=\columnwidth]{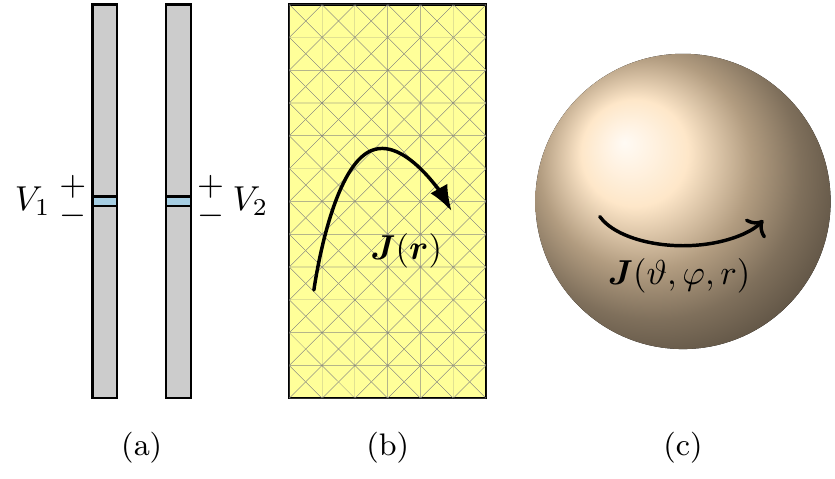}
\caption{Various source quantities~$\xv$ to be optimized. (a) Port voltage, $\xv = \Vv$, of two in-parallel placed dipoles. The arrangement embodies a~$\T{C}_\T{s}$ symmetry group. (b) Current density, $\xv = \Iv$, on a rectangular plate  belonging to a~$\T{C}_\T{2v}$ group. (c) Current density,~$\xv = \V{\alpha}$ on a spherical shell expanded in spherical harmonics. The spherical shell belongs to the~$\T{O}(3)$ symmetry group.}
\label{fig:fig1}
\end{figure}

Let us start with QCQP problem~$\OP{P}_1$ containing only quadratic terms
\begin{equation}
\begin{split}
\T{minimize} \quad & \xv^\herm \M{A} \xv \\
\T{subject\, to} \quad & \xv^\herm \M{B} \xv = 1 \\
& \xv^\herm \M{C} \xv = 0,
\end{split}
\label{eq:problem1}
\end{equation}
where $\xv$ is the optimized quantity, \eg{}, current density or port voltages in the source region, see Fig.~\ref{fig:fig1}, and~$\M{A}$,~$\M{B}$, and~$\M{C}$, are integro-differential operators represented in a basis defined by either piece-wise or entire-domain basis functions~$\left\{\basisFcn_n\right\}$, \cite{PetersonRayMittra_ComputationalMethodsForElectromagnetics}. It is assumed that the algebraic properties of the operators are compatible with the problem to be solved, \eg, $\M{B}\succ \M{0}$ and $\M{C}$ being generally indefinite (the true physical meaning of these operators is given later on).

The Lagrangian of the problem~$\OP{P}_1$ is
\begin{equation}
\OP{L}_1\left(\lambda_i, \xv\right) = \xv^\herm \M{H} \left(\lambda_i\right) \xv + \lambda_1
\label{eq:lagrang11}
\end{equation}
with its derivative
\begin{equation}
\dfrac{\partial \OP{L}_1\left(\lambda_i, \xv\right)}{\partial \xv^\herm} = \M{H} \left(\lambda_i\right) \xv,
\label{eq:lagrang12}
\end{equation}
where $\M{H} \left(\lambda_i\right) = \partial^2\OP{L}_1/\partial{\xv}^\herm\partial\xv = \M{A} - \lambda_1 \M{B} - \lambda_2 \M{C}$ is the Hessian matrix~\cite{NocedalWright_NumericalOptimization}. The stationary points~$\tilde{\xv}$ are solutions to
\begin{equation}
\dfrac{\partial \OP{L}_1\left(\lambda_i, \xv\right)}{\partial \xv^\herm} = \V{0},
\label{eq:lagrang13}
\end{equation}
or explicitly to
\begin{equation}
\M{A}\tilde{\xv} - \V{\lambda}_2 \M{C} \tilde{\xv} = \V{\lambda}_1 \M{B} \tilde{\xv}.
\label{eq:lagrang14}
\end{equation}

\subsection{Optimization Problem~$\OP{P}_2$}
\label{subsec:QCQP2}

For the sake of completeness, the second optimization problem~$\OP{P}_2$ is defined as
\begin{equation}
\begin{split}
\T{minimize} \quad & \xv^\herm \M{A} \xv \\
\T{subject\, to} \quad & \xv^\herm \M{B} \xv = 1 \\
& \xv^\herm \M{C} \xv = \RE\left\{\xv^\herm \M{b}\right\},
\end{split}
\label{eq:problem2}
\end{equation}
\ie{}, the second constraint contains a linear term in~$\xv$. Analogous to~\eqref{eq:lagrang11}, the Lagrangian reads
\begin{equation}
\OP{L}_2\left(\lambda_i, \xv\right) =
\xv^\herm\M{H} \left(\lambda_i\right)\xv + \lambda_2  \RE\left\{\xv^\herm \M{b}\right\} + \lambda_1
\label{eq:lagrang21}
\end{equation}
The derivative of the Lagrangian is
\begin{equation}
\dfrac{\OP{L}_2\left(\lambda_i, \xv\right)}{\partial \xv^\herm} = \M{H} \left(\lambda_i\right) \xv + \dfrac{\lambda_1}{2} \M{b}.
\label{eq:lagrang22}
\end{equation}
The stationary points~$\tilde{\xv}$ are
\begin{equation}
\tilde{\xv} = - \dfrac{\lambda_1}{2} \M{H}^{-1} \left(\lambda_i\right) \M{b},
\label{eq:lagrang23}
\end{equation}
with the demand that~$\M{H} \left(\lambda_i\right)\succ \M{0}$.

\subsection{Solution to Dual Problems}
\label{subsec:QCQPdual}

Primal problems~$\OP{P}_1$ and $\OP{P}_2$ with stationary points~\eqref{eq:lagrang14} and~\eqref{eq:lagrang23} are generally non-convex and are often approached using dual function~\cite{BoydVandenberghe_ConvexOptimization} defined as
\begin{equation}
d_p \left(\lambda_i\right) = \inf \limits_{\tilde{\xv}} \left\{ \OP{L}_p \left(\lambda_i, \tilde{\xv} \right)\right\},
\label{eq:dual1}
\end{equation}
where $p=\left\{1,2\right\}$. The supremum of the dual function
\begin{equation}
d_p^\ast = \sup_{\lambda_i} \left\{ d_p\left(\lambda_i\right) \right\},
\label{eq:dual2}
\end{equation}
is a lower bound to the primal optimization problem~\cite{BoydVandenberghe_ConvexOptimization}, the solution to which is here denoted as~$p^\ast$. Since the dual function is convex~\cite{BoydVandenberghe_ConvexOptimization}, the solution to~\eqref{eq:dual2} can easily be found. Algebraic techniques reducing the computational burden behind the optimization of this type of problem are presented in~\cite{2020_Gustafsson_NJP}. 

Generally, the duality gap~$g^\ast \geq 0$,
\begin{equation}
g^\ast = p^\ast - d^\ast
\label{eq:dual3}
\end{equation}
exists, nevertheless, problems involving the minimum Q-factor~\cite{CapekGustafssonSchab_MinimizationOfAntennaQualityFactor}, maximum antenna gain~\cite{GustafssonCapek_MaximumGainEffAreaAndDirectivity}, maximum radiation efficiency~\cite{UzsokySolymar_TheoryOfSuperDirectiveLinearArrays, Harrington_AntennaExcitationForMaximumGain}, minimum total active reflection coefficient (TARC)~\cite{CapekJelinekMasek_OptimalityOfTARC_Arxiv}, and their mutual trade-offs~\cite{GustafssonCapekSchab_TradeOffBetweenAntennaEfficiencyAndQfactor}, were shown to have no duality gap. Hence, to simplify the exposition, and without loss of generality, it is assumed for the rest of the paper that there is no duality gap~$g^\ast$, \ie{}, $p^\ast = d^\ast$ for both problems~$\OP{P}_1$ and $\OP{P}_2$.

The typical workflow solving problem~$\OP{P}_1$ consists of an iterative evaluation of the generalized eigenvalue problem~\eqref{eq:lagrang14}, taking the dominant eigenvalue~$\lambda_1$ and setting the multiplier~$\lambda_2$ so that~$\lambda_1$ is maximized. On the contrary, the treatment of problem~$\OP{P}_2$ requires a repetitive solution to the system of linear equations. This is an important distinction between problems~$\OP{P}_1$ and~$\OP{P}_2$: \textit{issues with symmetries may occur in problem~$\OP{P}_1$ while they cannot appear for problems of type~$\OP{P}_2$}.

\section{Illustrative Example -- Problem of $\OP{P}_1$-Type}
\label{subsec:PTexample}

Let us demonstrate the effect of symmetries on a practical example of Q-factor minimization with a constraint on the self-resonance of the current, specifically
\begin{equation}
\begin{split}
\T{minimize} \quad & \Iv^\herm \Wm \Iv \\
\T{subject\, to} \quad & \Iv^\herm \Rmvac \Iv = \dfrac{1}{2} \\
& \Iv^\herm \Xmvac \Iv = 0,
\end{split}
\label{eq:Qproblem1}
\end{equation}
where $\Wm = \M{A} = \omega\partial\Xmvac/\partial\omega$, $\Rmvac = \M{B}$, and $\Xmvac = \M{C}$ from~\eqref{eq:problem1}, \ie{}, the problem of the minimum Q-factor falls into a class of $\OP{P}_1$~problems, and $\M{Z}_0 = \Rmvac + \J \Xmvac$ is the impedance matrix for a scatterer made of a perfect electric conductor (PEC), see~\cite{2020_Gustafsson_NJP} for the exact definition of all the matrix operators. The basis functions used are RWG functions~\cite{RaoWiltonGlisson_ElectromagneticScatteringBySurfacesOfArbitraryShape} and the optimization variable~$\M{x} = \Iv$ represents the surface current density as
\begin{equation}
\Jv\left(\rv\right) \approx \sum_n I_n \basisFcn_n\left(\rv\right).
\label{eq:RWG}
\end{equation}
All the operators were evaluated in the AToM package~\cite{atom}.

This problem has a long history starting with a seminal work of Chu~\cite{Chu_PhysicalLimitationsOfOmniDirectAntennas} and has fully been described and solved in~\cite{CapekGustafssonSchab_MinimizationOfAntennaQualityFactor}. The solution to the dual problem~\eqref{eq:dual2} reads
\begin{equation}
d^\ast = \max_{\lambda_2} \min_m \lambda_{1,m}
\label{eq:Qsolution1}
\end{equation}
with the eigenvalues~$\lambda_{1,m}$ defined by
\begin{equation}
\dfrac{1}{2}\left( \Wm - \lambda_2 \Xmvac \right) \Iv_m = \lambda_{1,m} \Rmvac \Iv_m,
\label{eq:Qsolution2}
\end{equation} 
\cf{},~\eqref{eq:lagrang14}.

The definition of the Q-factor~\cite{IEEEStd_antennas} can be rewritten as~\cite{CapekGustafssonSchab_MinimizationOfAntennaQualityFactor}
\begin{equation}
Q\left(\Iv\right) = \dfrac{\max\left\{ \Iv^\herm \XMm \Iv, \Iv^\herm \XEm \Iv \right\}}{\Iv^\herm \Rmvac \Iv},
\label{eq:Qsolution3}
\end{equation}
where
\begin{subequations}
\begin{align}
\label{eq:Qsolution3BA}
\XMm  &= \dfrac{1}{2} \left( \Wm + \Xm \right), \\
\label{eq:Qsolution3BB}
\XEm  &= \dfrac{1}{2} \left( \Wm - \Xm \right).
\end{align}
\end{subequations}
The formula~\eqref{eq:Qsolution3} is valid for arbitrary current~$\Iv$ and can be used as a useful check of the duality gap~$g = Q(\lambda_2^\ast)-d^\ast$, where~$Q(\lambda_2^\ast)$ is a Q-factor evaluated via~\eqref{eq:Qsolution3} with current $\Iv_1$ ($m=1$) found by~\eqref{eq:Qsolution2} at $\lambda_2 = \lambda_2^\ast$. When no duality gap occurs, we have
\begin{equation}
Q^\ast = Q\left(\Iv_\T{opt}\right) = d^\ast.
\label{eq:Qsolution4}
\end{equation}

\begin{figure}
\centering
\includegraphics[]{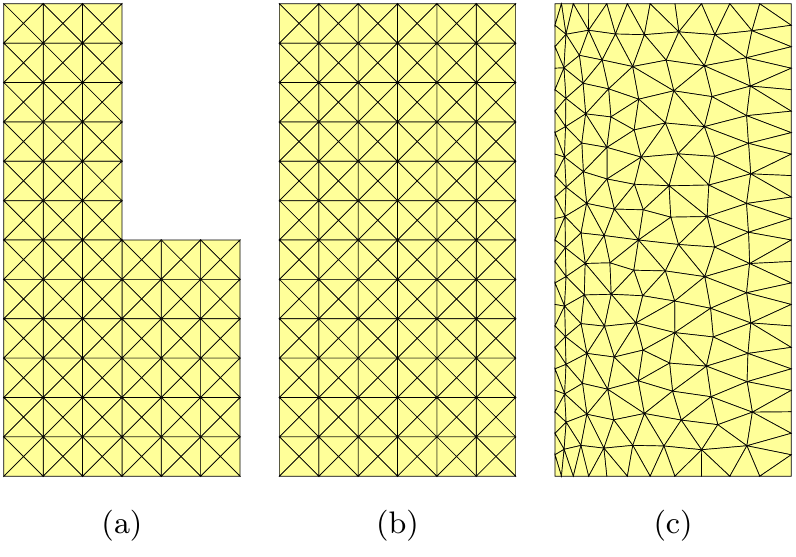}
\caption{Shapes and their discretization utilized to solve the optimization problem~\eqref{eq:Qproblem1}. (a) L-shape plate of dimensions~$\ell\times\ell/2$ with cutoff of size $\ell/2\times\ell/4$ discretized via Delaunay triangulation with a pixelized pattern consisting of $216$~triangles and $306$~basis functions. (b) The rectangular plate of dimensions~$\ell\times\ell/2$ with a mesh grid respecting the symmetries of the object consisting of $288$~triangles and $414$~basis functions. (c) Same as (b) with a non-symmetrical mesh grid intentionally made an-isotropic, consisting of $274$~triangles and $378$~basis functions. The electrical size is, in all cases, $ka = 1/2$, where $k$ is the wave-number and $a$ is the radius of the smallest sphere circumscribing the structure. All structures are made of PEC and the numerical quadrature of the third order~\cite{Dunavant_HighDegreeEfficientGQR} in AToM~\cite{atom} is utilized to gather the matrix operators.}
\label{fig:fig2}
\end{figure}

\begin{figure}
\centering
\includegraphics[width=\columnwidth]{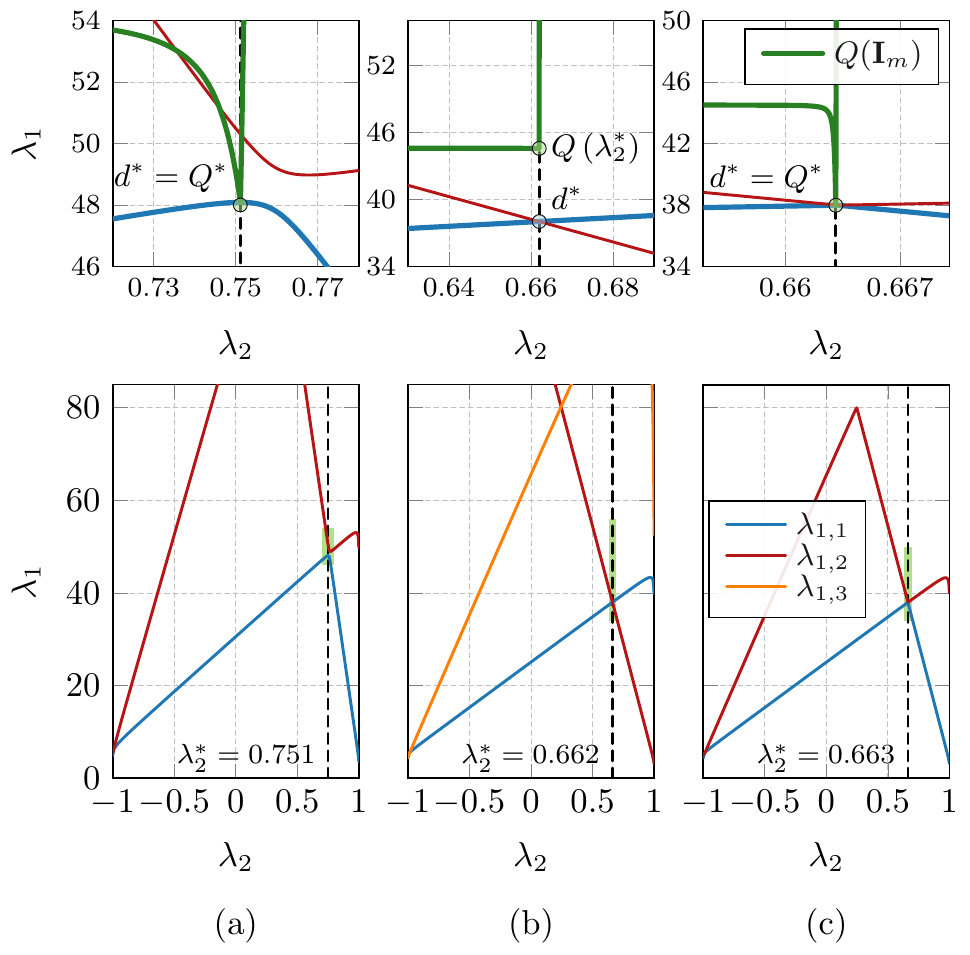}
\caption{Eigenvalues (Lagrange's multipliers)~$\lambda_1$ from~\eqref{eq:Qsolution2} as functions of Lagrange's multiplier~$\lambda_2$ (bottom panes) and the corresponding Q-factors~\eqref{eq:Qsolution3} (top panes, the green solid curves). The top panes show details in the vicinity of the optimal value of the multiplicator~$\lambda_2^\ast$. The two lowest eigenvalues~$\lambda_{1,m}$ are depicted. For case~(b), the lowest eigenvalue from $\textsc{B}_2$ (the blue line), $\textsc{A}_2$ (the red line), and $\textsc{B}_1$ (the orange line) irreducible representations~\cite{McWeeny_GroupTheory} are shown (to be discussed in the next section). The structures depicted in Fig.~\ref{fig:fig2} are employed with the physical setting described in the caption of Fig.~\ref{fig:fig2}.}
\label{fig:fig3}
\end{figure}

A solution to~\eqref{eq:Qsolution1} is found here for two different shapes: an L-shape plate and a rectangular plate with a perfectly symmetric mesh grid, see Fig.~\ref{fig:fig2}b. The effects of the non-regular mesh grid, depicted in Fig.~\ref{fig:fig2}c are studied as well. The dual function and its maximum~$d^\ast$ at $\lambda_2^\ast$ is shown in Fig.~\ref{fig:fig3} with subfigures (a)--(c) corresponding to those of Fig.~\ref{fig:fig2}. Due to the large numerical dynamics in the bottom panes, the vicinity of the dual solutions are zoomed in the top panes of Fig.~\ref{fig:fig3} with the traces for the actual value of Q-factor~\eqref{eq:Qsolution3} added. 

The non-symmetrical case (a) causes no problems and~$Q^\ast = d^\ast$ for $\lambda_2^\ast$, \ie{}, there is no duality gap. On the other hand, case~(b) seemingly embodies a duality gap~$g = Q\left(\lambda_2^\ast\right) - d^\ast$. This \Quot{erroneous} duality gap is caused by the eigenvalue crossing (two eigensolutions to~\eqref{eq:Qsolution2} are degenerate at~$\lambda_2^\ast$). Neither of the degenerated eigenmodes satisfy the last constraint of~\eqref{eq:Qproblem1}, which is manifested by the immediate increase in the value of corresponding Q-factor~$Q\left(\Iv_m\right)$. It is shown later on that the degenerate solutions to~\eqref{eq:Qsolution2} must properly be combined to satisfy this last constraint (to secure the self-resonance of the optimal current) and to close the gap. The last case~(c) has no duality gap thanks to the slightly non-symmetrical mesh grid.

This introductory example raises a series of questions:
\begin{enumerate}
    \item When can problems with symmetries be expected?
    \item How can the problem be detected?
    \item How can the erroneous duality gap~$g$ caused by the presence of symmetries be fixed?
    \item How can the robustness of the treatment for a numerical evaluation be improved?
\end{enumerate}
These questions are addressed in the following text after a brief review of the elements of point group theory.

\section{Presence of Symmetries}
\label{sec:Symmetries}

Point group theory\footnote{Only the crucial parts essential for this work are reviewed here. The reader is referred to, \eg{},~\cite{McWeeny_GroupTheory} and references therein for a comprehensive explanation.} constitutes the framework, both for the theoretical understanding and practical treatment of the issues related to symmetries.

Let us assume an object~$\srcRegion$ invariant to a set of point symmetries (\eg{}, rotation, reflection, etc.). Imagine further that object~$\srcRegion$ is discretized and basis functions~$\left\{\basisFcn_n\left(\V{r}\right)\right\}$ are applied. It can be shown~\cite{McWeeny_GroupTheory} that any operator, say~$\M{A}$, represented in a basis~$\left\{\basisFcn_n\left(\V{r}\right)\right\}$ and preserving the symmetries, can be block-diagonalized as
\begin{equation}
\widehat{\M{A}} = \Gm^\trans \M{A} \Gm = \begin{bmatrix}
\M{A}_1 & \cdots & \M{0} \\
\vdots & \ddots & \vdots \\
\M{0} & \cdots & \M{A}_G,
\end{bmatrix}
\label{eq:irrepsAssembly}
\end{equation}
where matrix~$\Gm$ is called a symmetry-adapted basis~\cite{McWeeny_GroupTheory} and its construction for piece-wise basis functions is shown, \eg{}, in~\cite{Maseketal_ModalTrackingBasedOnGroupTheory}. Each block~$\M{A}_g$ in~\eqref{eq:irrepsAssembly} belongs to a unique irreducible representation of the point group~\cite{McWeeny_GroupTheory}, briefly denoted hereinafter as \Quot{irreps}, see Appendix~\ref{sec:charTables} for some notable examples relevant to this work.

An important consequence of relation~\eqref{eq:irrepsAssembly} is that the eigenvalue decomposition of operator~$\M{A}$ on a symmetrical structure is also separable into irreps, \ie{}, each eigenvector belongs to a particular irrep and eigenvectors from different irreps are orthogonal to each each other even with respect to any operator. A central observation pertaining to the spectrum of the operator~$\widehat{\M{A}}$, attributed to von Neumann and Wigner~\cite{vonNeumannWigner_OnTheBehaviourOfEigenvaluesENG}, then states that if operator~$\M{A}$ is dependent on a certain parameter, such as frequency or Lagrange's multiplier, see~\eqref{eq:Qsolution2}, the traces of eigenvalues (abbreviated in this paper as \Quot{eigentraces}) belonging to the same irreps cannot cross each other~\cite{SchabEtAl_EigenvalueCrossingAvoidanceInCM,Maseketal_ModalTrackingBasedOnGroupTheory}. Applying this theorem to Fig.~\ref{fig:fig3}b, the blue and red traces must belong to modes from different irreps. Applying this theorem to Fig.~\ref{fig:fig3}a, no problems with degeneracies occur, since no symmetries are present, \ie{}, all modes belong to only one irrep, see Table~\ref{tab:charTableC1} of Appendix~\ref{sec:charTables}.

An useful side-product of~\eqref{eq:irrepsAssembly} is the acceleration of optimization since only the dominant solution from irrep~$\T{A}_2$ is needed for $\lambda_2 > \lambda_2^\ast$ and only the dominant solution from irrep~$\T{B}_2$ is needed for $\lambda_2 < \lambda_2^\ast$. That means a speed up by factor $4^q$ for point group $\T{C}_\T{2v}$ where the complexity of the algorithm is assumed~$\OP{O}(N^q)$. An extreme case is shown in~\cite{Knorr_1973_TCM_symmetry} for the body of revolution code where the system of basis functions forms a reducible system so that the matrix inverse is directly possible.

\section{Various Aspects of the Symmetry Presence}
\label{sec:examples}

Several problems of various complexity are solved and interpreted in this section in terms of point group theory. The necessity of combining two modes from different irreps to remove the erroneous duality gap is shown in Sec.~\ref{subsec:introproblem}. When geometry multiplicities appear, more than one solution exists and modes can freely be combined as shown in~Sec.~\ref{subsec:geom}. The study of how an imperfect mesh grid affects the symmetry treatment is conducted in Sec.~\ref{subsec:mesh}. Since the mesh grid is often made from rectangular or triangular elements, not all objects are perfectly represented, \eg{}, a spherical shell with triangular discretization elements~\cite{CapekEtAl_ValidatingCMsolvers}. The subsequent example in Sec.~\ref{subsec:portModes} shows that the theory introduced in this paper is generally valid for the arbitrary representation of the unknown (source) quantities, \cf{}, Fig.~\ref{fig:fig1}, by employing port-mode representation~\cite{1978_Harrington_TAP} to minimize the total active reflection coefficient~\cite{2005_ManteghiRahmatSamii_TARC, CapekJelinekMasek_OptimalityOfTARC_Arxiv} of the metallic rim. The spectrum of the spherical shell is evaluated analytically and compared with the numerical solution in Sec.~\ref{subsec:sphere}. The last example in Sec.~\ref{subsec:modeManipul} deals with an academic, yet highly relevant, technique manipulating the eigenvalue traces of the isolated irrep.

\subsection{Algebraic Multiplicity of Eigenvalues (Rectangular Plate)}
\label{subsec:introproblem}

The erroneous duality gap shown in Sec.~\ref{subsec:PTexample} for a rectangular plate, see Fig.~\ref{fig:fig2}b and the results in Fig.~\ref{fig:fig3}b, is eliminated here by the proper combination of degenerate eigenvectors. 

\begin{figure}
\centering
\includegraphics[]{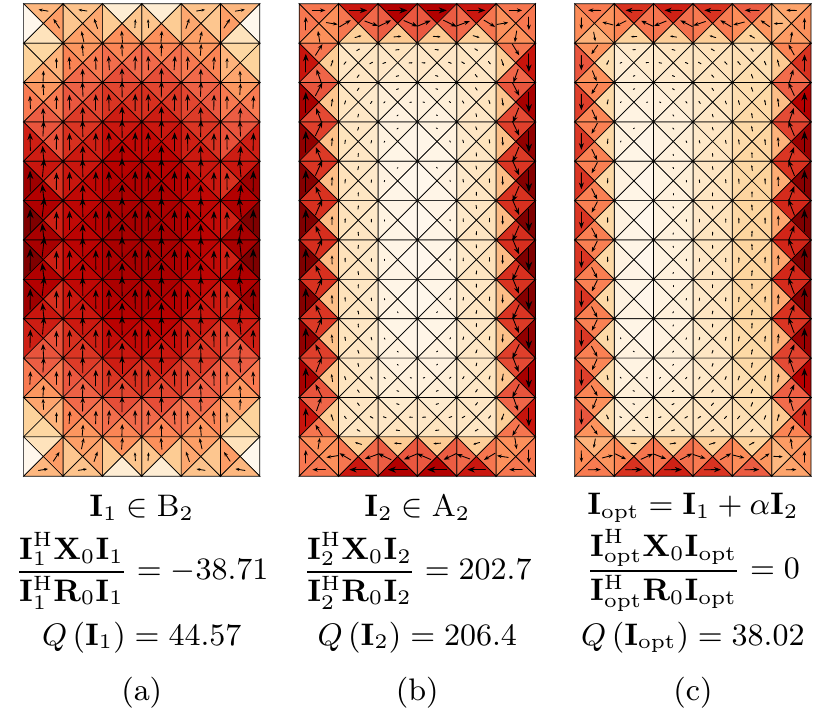}
\caption{Current densities associated with the first two modes of the eigenvalue problem~\eqref{eq:Qsolution2} evaluated for the rectangular shape depicted in Fig.~\ref{fig:fig2}b at $\lambda_2 = \lambda_2^\ast$, \cf{}~Fig.~\ref{fig:fig3}b. The right subfigure shows the correct combination to eliminate the erroneous duality gap depicted in the top pane of Fig.~\ref{fig:fig3}b. Subfigure (a) shows a capacitive mode belonging to irrep~$\T{B}_2$ (the blue line in Fig.~\ref{fig:fig3}b) with Q-factor~$Q\left(\Iv_1\right) = 44.57$. Subfigure (b) shows an inductive mode belonging to irrep~$\T{A}_2$ (the red line in Fig.~\ref{fig:fig3}b) with Q-factor~$Q\left(\Iv_1\right) = 206.4$. Finally, subfigure (c) shows the combination of currents from subfigures (a) and (b) with the mixing coefficient~$\alpha = 4.232$. The resulting current~$\Iv_\T{opt}$ is self-resonant and $Q\left(\Iv_\T{opt}\right) = Q^\ast = d^\ast$.}
\label{fig:fig5}
\end{figure}

The optimization problem~\eqref{eq:Qproblem1} is solved with~\eqref{eq:Qsolution2} by separately utilizing~\eqref{eq:irrepsAssembly} for irreps~$\T{B}_2$ and~$\T{A}_2$, \ie{}, two traces with a crossing at~$\lambda_2^\ast=0.662$ in Fig.~\ref{fig:fig3}b. At the crossing point, the corresponding eigenvectors can be linearly combined without a change of dual function value~$g^\ast = \lambda_1 \left(\lambda_2^\ast\right)$. Taking dominant modes from irreps as $\Iv_a \in \T{B}_2$, $\Iv_b \in \T{A}_2$, see Fig.~\ref{fig:fig5}, we get
\begin{equation}
\Iv_\T{opt} = \Iv_a + \alpha \Iv_b.
\label{eq:Iopt}
\end{equation}
The erroneous duality gap in Fig.~\ref{fig:fig3}b, top pane, is a manifestation of the constraint's violation in~\eqref{eq:Qproblem1}. Therefore, constant~$\alpha$ is found to fulfill
\begin{equation}
\Iv_\T{opt}^\herm \M{C} \Iv_\T{opt} = 0.
\label{eq:IoptCondition}
\end{equation}
Since modes~$\Iv_a$ and $\Iv_b$ belong to different irreps, we have $\Iv_a^\herm \M{C} \Iv_b = 0$, and
\begin{equation}
\alpha = \sqrt{-\dfrac{\Iv_a^\herm \M{C} \Iv_a}{\Iv_b^\herm \M{C} \Iv_b}} \T{e}^{\J \varphi}
\label{eq:IoptSolution}
\end{equation}
with~$\varphi\in[0, 2\pi)$ and the assumption that the square root is real. Combining the degenerated modes with~\eqref{eq:Iopt}, \eqref{eq:IoptSolution}, we get the optimal current, see Fig.~\ref{fig:fig5}c, fulfilling all constraints and~$d^\ast = Q^\ast$.

Notice that the mixing coefficient~$\alpha$ has the same form as in~\cite{CapekJelinek_OptimalCompositionOfModalCurrentsQ}, where two dominant characteristic modes (capacitive and inductive) were combined to get a minimum Q-factor.

\subsection{Geometry Multiplicity of Eigenvalues (Square Plate)}
\label{subsec:geom}

This example attempts to highlight the difference between a degeneracy across irreps (the previous section) and the higher dimension of a single irrep, a situation where the symmetries introduce additional degrees of freedom (this section).

\begin{figure}
\centering
\includegraphics[]{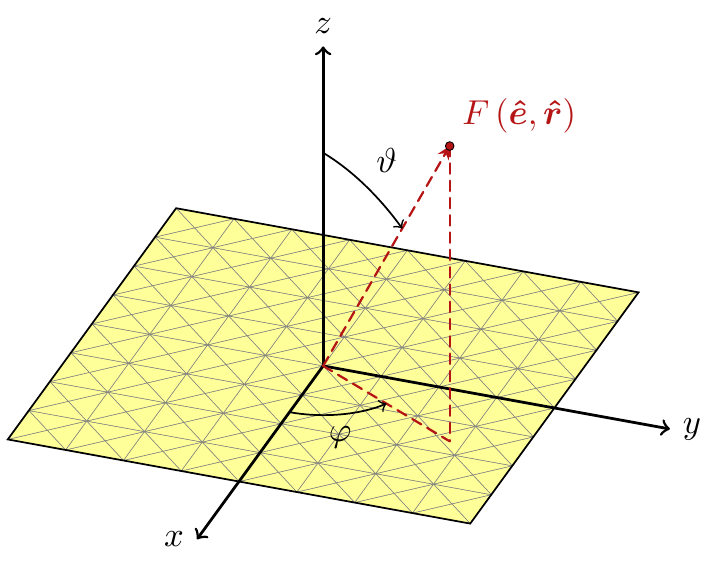}
\caption{Optimization setting and coordinate system used for the optimization of maximal antenna gain with a self-resonant constraint. Symbol~$F\left(\UV{e},\UV{r}\right)$ denotes an electric far field in $\UV{r}$ direction and of polarization~$\UV{e}$.}
\label{fig:fig8}
\end{figure}

Let us consider a setting depicted in Fig.~\ref{fig:fig8} which shows a square plate made of a perfectly conducting material, discretized with a symmetric mesh grid, and centered with respect to the coordinate system. Antenna gain~$G$ in a direction~$\UV{r}$ and polarization~$\UV{e}$ for a self-resonant current is to be maximized as
\begin{equation}
\begin{split}
\T{minimize} \quad & - \Iv^\herm \Um \left( \UV{e}, \UV{r} \right) \Iv \\
\T{subject\, to} \quad & \Iv^\herm \left(\Rmvac + \Rmmat \right) \Iv = 1 \\
& \Iv^\herm \Xmvac \Iv = 0,
\end{split}
\label{eq:Gproblem3}
\end{equation}
where~$\M{U}$ is a radiation intensity matrix with low-rank representation~\cite{JelinekCapek_OptimalCurrentsOnArbitrarilyShapedSurfaces}
\begin{equation}
\Um\left( \UV{e}, \UV{r} \right) = \Fv^\herm \left( \UV{e}, \UV{r} \right) \Fv \left( \UV{e}, \UV{r} \right),
\label{eq:Gproblem4}
\end{equation}
where $\Fv \left( \UV{e}, \UV{r} \right) = \begin{bmatrix} \Fv_{\UV{\vartheta}}^\trans \left( \UV{r} \right) & \Fv_{\UV{\varphi}}^\trans \left(\UV{r} \right) \end{bmatrix}^\trans$ and $\Rmmat$ is a material matrix defined in~\cite{JelinekCapek_OptimalCurrentsOnArbitrarilyShapedSurfaces}. The optimization problem~\eqref{eq:Gproblem3} is solved according to the procedure from Sec.~\ref{subsec:QCQP1} by combining constraints as proposed in~\cite{GustafssonCapek_MaximumGainEffAreaAndDirectivity}. The solution reads
\begin{equation}
G^\ast = d^\ast = -4\pi\max_{\nu} \min_m \lambda_{1,m},
\label{eq:Gsolution1}
\end{equation}
where
\begin{equation}
- \Um \Iv_m  = \lambda_{1,m} \left( \Rmvac + \Rmmat - \nu \Xmvac \right) \Iv_m
\label{eq:Gsolution2}
\end{equation}
with~$\nu   = - \lambda_2/\lambda_{1,m} \in [\nu_\T{min}, \nu_\T{max}]$ being picked so that the matrix on the right-hand side of~\eqref{eq:Gsolution2} is positive definite~\cite{GustafssonCapek_MaximumGainEffAreaAndDirectivity}. A further acceleration of the formula~\eqref{eq:Gsolution2} is possible, see~\cite{GustafssonCapek_MaximumGainEffAreaAndDirectivity} for details.

The optimization problem~\eqref{eq:Gproblem3} differs from~\eqref{eq:Qproblem1} in two respects. First, matrix~$\Um$ has rank 2, which means that only two eigenvalues from~\eqref{eq:Gsolution2} differ from zero. Second, matrix~$\Um$ explicitly depends on the the observation coordinate, which also must be taken into account when considering the symmetries of the problem. Notice that for a general observation coordinate~$\UV{r}$, the physical problem is not symmetric although the antenna geometry is.

For the purpose of this example, let us assume that the direction for radiation intensity maximization has been set to~$\UV{r} = \UV{z}$ and that the electrical size is~$ka = 1/2$. The material parameters were set to be equivalent to copper at frequency~$f = 1\,$GHz. No restrictions were imposed on polarization~$\UV{e}$ meaning that the solution can equally be formed by polarization pointing into $\UV{\vartheta}$ and $\UV{\varphi}$ directions (or their combination). With these settings and the mesh grid from Fig.~\ref{fig:fig8}, the optimization problem complies with symmetries of the~$\T{C}_{4\T{v}}$ point group, see Table~\ref{tab:charTableC4v} in Appendix~\ref{sec:charTables}. 

\begin{figure}
\centering
\includegraphics[width=\columnwidth]{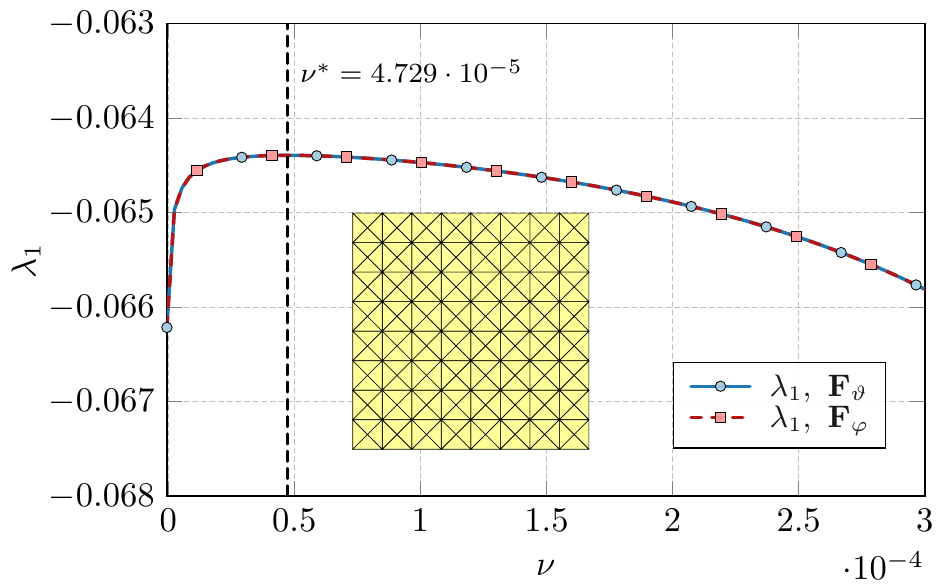}
\caption{Solution to the dual problem~\eqref{eq:Gsolution2} for a rectangular plate of electrical size~$ka = 1/2$ made of lossy material equivalent to copper at $1$\,GHz. The observation direction is~$\UV{r} = \UV{z}$. The inset shows the mesh grid utilized for the optimization. The optimum value of Lagrange's multiplier~$\nu^\ast = 4.729\cdot 10^{-5}$ is highlighted by the dashed black line. The two depicted eigentraces belong to irrep~$\T{E}$, see Table~\ref{tab:charTableC4v} in Appendix~\ref{sec:charTables}, and are twice degenerated for all values of~$\nu$.}
\label{fig:fig06}
\end{figure}

The solution to~\eqref{eq:Gsolution2} is depicted in Fig.~\ref{fig:fig06} with an immediate observation of twice degenerated eigentraces. These traces belong to irrep~$\T{E}$ (the only two-dimensional irrep of point group~$\T{C}_{4\T{v}}$). Since there is no other eigentrace crossing these two at~$\nu^\ast$ (all other eigenvalues are zero), there is no need to combine modes to fulfill the third constraint as in Sec.~\ref{subsec:introproblem}. Instead, both solutions are valid on their own. They are geometry multiplicities, because for~$\UV{r} = \UV{z}$ the two rank-one matrices $\Fv_{\UV{\vartheta}}$ and $\Fv_{\UV{\varphi}}$ forming operator~$\Um$ are linearly dependent
\begin{equation}
\Fv_{\UV{\varphi}} = \Fv_{\UV{\vartheta}} \M{C}_4,
\label{eq:Girrep2}
\end{equation}
where~$\M{C}_4 \in \mathbb{R}^{N\times N}$ is the (unitary) rotation matrix by angle~$\varphi = \pi/2$ around $\UV{z}$ axis, $\M{C}_4^\herm \M{C}_4 = \M{1}$, represented in basis~$\left\{\basisFcn_n\left(\V{r}\right)\right\}$, therefore,
\begin{equation}
\Fv = \begin{bmatrix} \Fv_{\UV{\vartheta}} \\ \Fv_{\UV{\vartheta}} \M{C}_4 \end{bmatrix}
\label{eq:Girrep1}
\end{equation}
yields twice degenerated eigenvalue~$\lambda_1$ in~\eqref{eq:Gsolution2} since according to~\eqref{eq:Gproblem4}
\begin{equation}
\Fv^\herm \Fv = \Fv_{\UV{\vartheta}}^\herm \Fv_{\UV{\vartheta}} + \M{C}_4^\herm \Fv_{\UV{\vartheta}}^\herm \Fv_{\UV{\vartheta}} \M{C}_4.
\label{eq:Girrep1B}
\end{equation}

Adapting the knowledge gained in this section on an example of minimal Q-factor optimization from the previous section with a shape from, \eg{}, the $\T{C}_{\T{4v}}$ point group (a square plate), a problem originates where two modes out of three degeneracies have to be combined as~\eqref{eq:Iopt} to fulfil the third constraint~\eqref{eq:IoptCondition}. In such a case, these two modes have to be from different irreps, specifically~$\Iv_a \in \OP{I}_A$, $\Iv_b \in \OP{I}_B$, $A \neq B$ so that
\begin{equation}
\T{sign}\left\{\Iv_a^\herm \Xmvac \Iv_a\right\} = - \T{sign}\left\{\Iv_b^\herm \Xmvac \Iv_b \right\},
\label{eq:Girrep3}
\end{equation}
otherwise the erroneous duality gap cannot be eliminated.

\subsection{Imperfections of the Mesh Grid}
\label{subsec:mesh}

The understanding gained in the previous sections will be exploited here on an example of mesh grid imperfectness, where the point group rules are obscured by fact that all computations are made with finite numerical precision.

\begin{figure}
\centering
\includegraphics[width=\columnwidth]{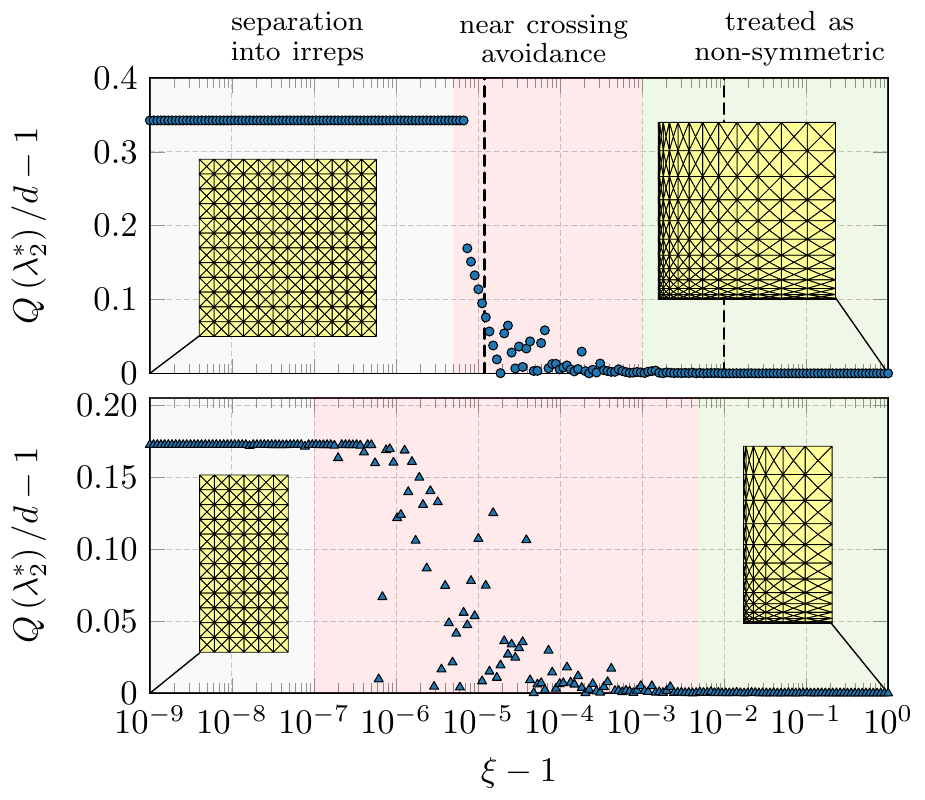}
\caption{Study of an erroneous duality gap expressed in terms of~$Q\left(\lambda_2^\ast\right)$ for an optimization of self-resonant Q-factor. Two symmetric objects are considered, a square plate ($\T{C}_\T{4v}$~point group, top pane) and a rectangular plate ($\T{C}_\T{2v}$~points group, bottom pane). The parameter~$\xi$ distorts the mesh grid both in horizontal and in vertical directions, see the insets. The point group theory with separation into irreps applies for the negligible distortions, see left parts of the panes, highlighted by the gray background. For significant distortions, the structures behave as non-symmetric and no duality gap appears, see the right parts of the panes, highlighted by the green color. The most problematic part is the transition between symmetric and non-symmetric cases, see the intermediate parts of the panes, highlighted by the red color. The dashed black lines corresponds to subfigures (b) and (c) in Fig.~\ref{fig:fig7}.}
\label{fig:fig4}
\end{figure}

Two structures of different point groups are assumed, a square plate ($\T{C}_{4\T{v}}$) and a rectangular plate ($\T{C}_{2\T{v}}$). The optimization of the Q-factor introduced in Sec.~\ref{sec:Symmetries} and solved in Sec.~\ref{subsec:introproblem} is considered. The discretization grids are made of square pixels, see the insets on the left of Fig.~\ref{fig:fig4}, or compressed both horizontally and vertically, see the insets on the right of Fig.~\ref{fig:fig4}. Assuming that the mesh grid lies in the $x-y$ plane with the bottom-left corner at the origin, the compression is provided via transformation
\begin{equation}
\left[\dfrac{x}{L_x}, \dfrac{y}{L_y} \right] \to \left[\left(\dfrac{x}{L_x}\right)^\xi, \left(\dfrac{y}{L_y}\right)^\xi \right],
\label{eq:contraction}
\end{equation}
applied on every grid node, where~$L_x,L_y$ are side lengths of the square or rectangle and~$\xi \in \left(1, \infty \right)$. 

The smooth distortion of the symmetric mesh grid enables an evolution of an erroneous duality gap to be seen, depicted as a normalized quantity in Fig.~\ref{fig:fig4}. For~$\xi = 1$, the mesh grids preserve the symmetry of the object and an erroneous duality gap exists, see the left part of Fig.~\ref{fig:fig4} highlighted by the gray background color. The error given by the difference between the primal and dual solution attains~$34\,\%$ for the square plate and about~$17\,\%$ for the rectangular plate, respectively. For a reasonable large value of~$\xi$, say $\xi > 1 + 10^{-2}$, the non-symmetry of the grid is significant enough that no special treatment is required (duality gap is zero), see the right part of Fig.~\ref{fig:fig4}, highlighted by the green background color. The most challenging cases lie between these two regions, highlighted by the red background color in Fig.~\ref{fig:fig4}, and often occur in practice due to rounding errors and other numerical imperfections. This region deserves further attention because the symmetry treatments from the previous sections have to be properly adapted.

\begin{figure}
\centering
\includegraphics[width=\columnwidth]{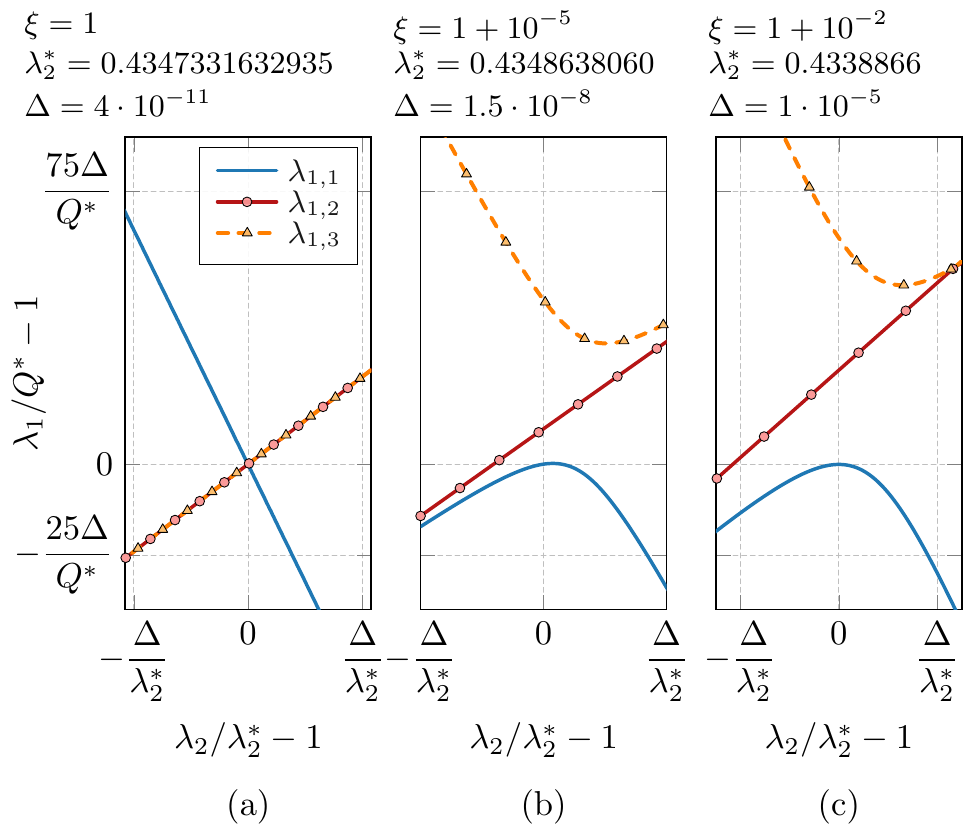}
\caption{Investigation of the close vicinity of the eigenvalue crossing/crossing avoidances for three particular cases from Fig.~\ref{fig:fig4}, top pane (the square plate). From left to right, they are evaluated for (a) $\xi = 1$ (symmetric mesh grid), (b) $\xi = 1 + 10^{-1}$ (slightly non-symmetrical mesh grid), and (c) $\xi = 1 + 10^{-2}$ (non-symmetric mesh grid). Due to the enormous sensitivity of the numerical precision, the high-order quadrature rule was applied to evaluate the matrix operators.}
\label{fig:fig7}
\end{figure}

The dual solution to the example of the square plate and Q-factor minimization, depicted in Fig.~\ref{fig:fig4}, top pane, is repeated in Fig.~\ref{fig:fig7}. The close vicinity around $\lambda_2^\ast$ point is studied for~$\xi = \left\{1, 1 + 10^{-5}, 1 + 10^{-2} \right\}$, \ie{}, for three various representatives of different regions in Fig.~\ref{fig:fig4}. It is seen that crossings of eigentraces for the symmetric case evolves into the crossing avoidance scenario initially described in~\cite{vonNeumannWigner_OnTheBehaviourOfEigenvaluesENG} (English transcription) and recalled in~\cite{SchabEtAl_EigenvalueCrossingAvoidanceInCM}. A problematic case appears in~Fig.~\ref{fig:fig7}b where the values of $\lambda_1$ for $\lambda_2^\ast$ are very close to each other, in this particular case they are the same up to six significant digits, yet not separated into irreps. The assumption~$\Iv_a^\herm \M{C} \Iv_b = 0$ secured by~\eqref{eq:irrepsAssembly} is, therefore, not valid anymore and one has to solve~\eqref{eq:IoptCondition} via~\eqref{eq:Iopt} using
\begin{equation}
|\alpha|^2 + 2 \dfrac{\RE \left\{\alpha \Iv_a^\herm \Xmvac \Iv_b \right\}}{\Iv_b^\herm \Xmvac \Iv_b} + \dfrac{\Iv_a^\herm \Xmvac \Iv_a}{\Iv_b^\herm \Xmvac \Iv_b} = 0.
\label{eq:IoptNum}
\end{equation}

Formula~\eqref{eq:IoptNum} is a generalization of~\eqref{eq:IoptSolution} for slightly non-symmetrical structures or for perturbed non-symmetrical mesh grids of symmetric structures. The only difference is the selection of suitable modes~$\Iv_a$ and $\Iv_b$ to be combined. This fact is discussed further on the algorithmic level in the next section. Notice that the scenario shown in Fig.~\ref{fig:fig7}c contains one eigentrace (the blue curve) which is significantly separated from the others to represent the true solution to the problem on its own.

\subsection{Change of Basis (TARC of a Lossy Metallic Rim)}
\label{subsec:portModes}

It is shown in this example that the presence of symmetries strongly affects the physics even when the basis~\eqref{eq:RWG} is changed, \ie{}, the operators are represented in another basis, which is still compatible with the point group of the studied object. A prominent example of this behaviour is a port modes representation~\cite{1978_Harrington_TAP}, which advantageously reduces the size of the problem. Another advantage is that since the unknowns are the terminal voltages, \cf{}, Fig.~\ref{fig:fig1}b, the optimal solution is directly realizable.

\begin{figure}
\centering
\includegraphics[]{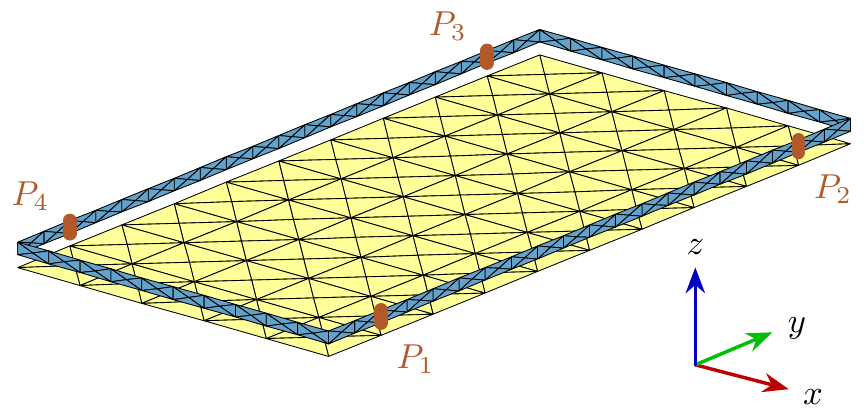}
\caption{A metallic rim with parasitic ground plane with four discrete ports, denoted $P_1,\dots,P_4$. Both the rim and the ground plane are made of copper. The ports are placed at the distance $\ell/5$ from the ends of the longer side.}
\label{fig:fig12}
\end{figure}

A metallic rim placed over parasitic ground plane is shown in Fig.~\ref{fig:fig12}. The size of the ground plane is~$150\,\T{mm} \times 75\,\T{mm}$, the height of the rim is~$2.5$\,mm and the height over the ground plane is~$2.5$\,mm (the dimensions are adjusted to be equivalent to a smart phone chassis). The material of the chassis is copper. The discretization grid was generated to accommodate the~$\T{C}_{2\T{v}}$ point group. The total active reflection coefficient (TARC),~\cite{2005_ManteghiRahmatSamii_TARC}, as defined for port mode quantities in~\cite{CapekJelinekMasek_OptimalityOfTARC_Arxiv} is to be optimized. The degrees of freedom are the terminal voltages, the characteristic impedance of the transmission line~$R_{0,i}$ and the matching susceptances~$B_{\T{L},i}$, see~\cite{CapekJelinekMasek_OptimalityOfTARC_Arxiv}~for the detailed optimization procedure.

The position of the ports is specified in Fig.~\ref{fig:fig12}, with the polarization of the delta gaps pointing towards $+\UV{y}$ direction. Port admittance matrix~$\M{y}$ is of $4 \times 4$~size and complies with the symmetries of the~$\T{C}_{2\T{v}}$ point group. The port voltages and admittances~$\left( R_{0,i}^{-1} - \J B_{\T{L},i} \right)$ enforcing simultaneously zero reflections on all ports are obtained as solutions to an eigenvalue problem~\cite{CapekJelinekMasek_OptimalityOfTARC_Arxiv}
\begin{equation}
\M{y}\M{v}_i = \left( R_{0,i}^{-1} - \J B_{\T{L},i} \right) \M{v}_i
\label{eq:minTARC}
\end{equation}
and are depicted in Table~\ref{tab:optimFeeding} one by one as belonging to different irreps. When properly normalized, they evoke the character table for the~$\T{C}_{2\T{v}}$~point group, see Table~\ref{tab:charTableC2v} in Appendix~\ref{sec:charTables}. If port~$P_1$ is taken as the initial port, port~$P_3$ is identified as its rotation by~$\pi$, port~$P_2$ as reflection through $xz$~plane, and port~$P_4$ as reflection through $yz$~plane. Knowing this, the voltage solutions can be assigned to the irreps they represent. 

The initial values of matching~$B_\T{L}$ and loading~$R_0$ given by~\eqref{eq:minTARC} can further be optimized as described in~\cite{CapekJelinekMasek_OptimalityOfTARC_Arxiv}. TARC values~$\Gamma^\T{t}$ for all excitation schemes are summarized in the last column of Table~\ref{tab:optimFeeding}, concluding that the feeding scheme~$\M{v}_3$, \ie{}, with the voltage orientation along the loop formed by the rim, gives the minimum~TARC. This excitation scheme belongs to the~$\T{B}_1$~irrep and dominates up to frequency~$f \approx 750$\,MHz. Around that frequency the best performing irrep switches to another one.

One notable implication of the symmetries is that the voltage schemes from~Table~\ref{tab:charTableC2v} are identical in amplitude which simplifies the feeding circuitry, see~\cite{Maseketal_ExcitationSchemesOfUncorrelatedChannels_Arxiv}~for a detailed study.

\begin{table}[]
\centering
\caption{Summary of TARC optimization for rectangular metallic rim with a parasitic ground plane for symmetry placement of four ports.}
\begin{tabular}{ccccccccc}
 & $P_1$ & $P_2$ & $P_3$ & $P_4$ & irrep & $1/B_\T{L}$ & $R_0$ & $\Gamma^\T{t}$ \\ \toprule
$\M{v}_1$ & $+1$ & $+1$ & $+1$ & $+1$ & $\T{A}_1$ & $451.3$ & $24800$ & $0.3064$ \\
$\M{v}_2$ & $+1$ & $-1$ & $+1$ & $-1$ & $\T{A}_2$ & $-313.8$ & $156700$ & $0.4212$ \\
$\M{v}_3$ & $+1$ & $+1$ & $-1$ & $-1$ & $\T{B}_1$ & $\mathbf{28.64}$ & $\mathbf{98.52}$ & $\mathbf{0.2374}$ \\
$\M{v}_4$ & $+1$ & $-1$ & $-1$ & $+1$ & $\T{B}_2$ & $21.23$ & $149.7$ & $0.3302$ \\ \bottomrule
\end{tabular}
\label{tab:optimFeeding}
\end{table}

The conclusions drawn in this section, \ie{}, that the effects of the symmetries remain the same with a proper change of basis, apply for many practical applications. For example, the entire domain basis of characteristic modes~\cite{HarringtonMautz_TheoryOfCharacteristicModesForConductingBodies} suffers from the necessity of eigentrace tracking~\cite{SchabEtAl_EigenvalueCrossingAvoidanceInCM, Maseketal_ModalTrackingBasedOnGroupTheory}. On the other hand, proper use of symmetries introduces additional degrees of freedom, \eg{}, for MIMO antenna design~\cite{Peitzmeier_Manteuffel-UpperBoundsForUncorrelatedPorts2019, MautzHarrington_ACombinedSourceSolutionForRadiationAndScatteringFromAPCbody}. Another notable example involves reduction with the Schur complement~\cite{CapekGustafssonSchab_MinimizationOfAntennaQualityFactor}.

\subsection{Analytically Solvable Problem (A Spherical Shell)}
\label{subsec:sphere}

The next optimization problem is solved analytically. A minimal dissipation factor~$\delta$ \cite{Harrington_AntennaExcitationForMaximumGain} is found with the optimal current being self-resonant~\cite{Jelinek+etal2018}. A spherical shell of radius~$a$ and electrical size~$ka$ is considered. Explicitly, the optimization problem reads~\cite{Jelinek+etal2018, GustafssonCapekSchab_TradeOffBetweenAntennaEfficiencyAndQfactor}
\begin{equation}
\begin{split}
\T{minimize} \quad & \Plost \\
\T{subject\, to} \quad & \Prad = 1 \\
& P_\T{react} = 0,
\end{split}
\label{eq:deltaProblem1}
\end{equation}
where the value of lost power~$\Plost$, radiated power~$\Prad$, and reactive power~$P_\T{react} = 2\omega\left( W_\T{m} - W_\T{e} \right)$ is given by quadratic forms as before. The optimal dissipation factor is evaluated as $\delta = \Plost/ \Prad$ \cite{Harrington_AntennaExcitationForMaximumGain}.

\begin{figure}
\centering
\includegraphics[width=\columnwidth]{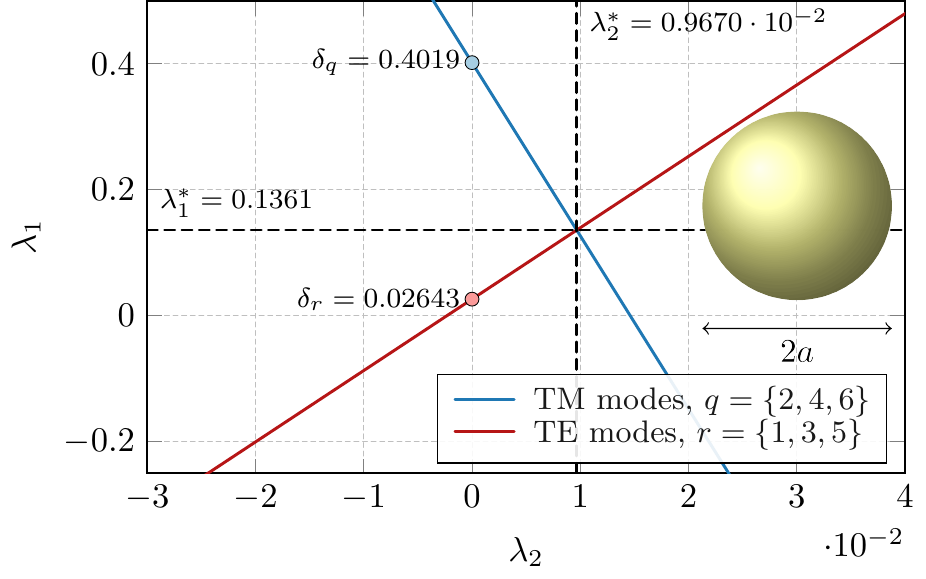}
\caption{Eigenvalue (Lagrange's multipliers) $\lambda_1$ from~\eqref{eq:deltaProblemN2} as functions of Lagrange's multiplier~$\lambda_2$. The TM modes are presented by the red eigentraces (since the characteristic number is negative, $\lambda_q < 0$, the curves are increasing). The TE modes are represented by the blue eigentraces (since~$\lambda_r > 0$). Both curves are three times degenerated (in correspondence with the geometrical multiplicity of dominant TM and TE modes).}
\label{fig:fig13}
\end{figure}

Let us start with a proper representation of the operators, here, in an entire domain basis of regular spherical waves~$\V{u}_p \left( k \V{r} \right)$~\cite{Hansen_SphericalNearFieldAntennaMeasurements}
\begin{equation}
\V{J} \left(\V{r}\right) = \sum_p \alpha_p \V{u}_p \left( k \V{r} \right).
\label{eq:deltaProblem1A}
\end{equation}
The operators are given element-wise as
\begin{equation}
R_{0, pq} = \langle \V{u}_p , \OP{R}_0 (\V{u}_q) \rangle = Z_0 k \int_\srcRegion \int_{\srcRegion '} U_{pq} \dfrac{\sin\left(k R \right)}{4 \pi R} \D{V} \D{V'},
\label{eq:deltaProblem2}
\end{equation}
with
\begin{equation}
U_{pq} = \V{u}_p^\ast \left( k \V{r} \right) \cdot \V{u}_q \left( k \V{r}' \right) - \dfrac{1}{k^2} \nabla \cdot \V{u}_p^\ast \left( k \V{r} \right) \nabla' \cdot \V{u}_q \left( k \V{r}' \right),
\label{eq:deltaProblem2B}
\end{equation}
$\Rmvac = [R_{0,pq}]$, and similarly for~$\Xm_0$ and~$\Rmmat$, see~\cite{JelinekCapek_OptimalCurrentsOnArbitrarilyShapedSurfaces}, where~$\M{u}_p \left( k \V{r} \right)$ is a regular spherical wave with the multi-index
\begin{equation}
p = 2 \left( l^2 + l - 1 + \left(-1\right)^s m \right) + \tau,
\label{eq:deltaProblem3}
\end{equation}
with all variables described, \eg{}, in~\cite{TayliEtAl_AccurateAndEfficientEvaluationofCMs}, $Z_0$ is impedance of a vacuum, $k$ is wavenumber, and $R = |\V{r} - \V{r}'|$. Importantly, the choice of spherical waves for spherical object leads to diagonal matrices~$\Rmmat$, $\Rmvac$, and~$\Xmvac$. Consequently, the eigenvalue problem~\eqref{eq:lagrang14} for the problem~\eqref{eq:deltaProblem1} reads
\begin{equation}
\left( \Rmmat - \lambda_2 \Xmvac \right) \V{\alpha} = \lambda_1 \Rmvac \V{\alpha}
\label{eq:deltaProblemN1}
\end{equation}
which can further be separated into individual equations for each spherical wave
\begin{equation}
\dfrac{R_{\rho p}}{R_{0 p}} - \lambda_2 \dfrac{X_{0 p}}{R_{0 p}} = \delta_p - \lambda_2 \lambda_p = \lambda_{1p},
\label{eq:deltaProblemN2}
\end{equation}
where~$\delta_p$ is the dissipation factor~\cite{Harrington_EffectsOfAntennaSizeOnGainBWandEfficiency}, and~$\lambda_p$ is a characteristic number\footnote{Spherical harmonics are the characteristic modes of a spherical shell~\cite{CapekEtAl_ValidatingCMsolvers}.}, both being evaluated for dominant spherical waves in~\cite{2017_Losenicky_Comment_Pfeiffer}. Since the dissipation factors~$\delta_p$ are positive and characteristic numbers~$\lambda_p$ are indefinite, \eqref{eq:deltaProblemN2} generates straight lines increasing (decreasing) with multiplicator~$\lambda_2$ for capacitive, $\lambda_p < 0$ (inductive, $\lambda_p > 0$) modes, see Fig.~\ref{fig:fig13}.

To solve the dual problem~\eqref{eq:dual2}, two modes, say the $q$th and the $r$th spherical waves, have to be chosen so that their traces intersect with the lowest value of~$\lambda_1$. This task is accomplished by taking the dominant TM and TE modes, $q \in \left\{2, 4, 6 \right\}$ and $r \in \left\{1, 3, 5 \right\}$, respectively, with
\begin{equation}
\lambda_2^\ast = \dfrac{\delta_r - \delta_q}{\lambda_r - \lambda_q}.
\label{eq:deltaProblemN3}
\end{equation}
Substituting~\eqref{eq:deltaProblemN3} into~\eqref{eq:deltaProblemN2} for $p=q$ or $p=r$ yields $\lambda_1^\ast = \delta^\ast$, see Fig.~\ref{fig:fig13}. The solution~\eqref{eq:deltaProblemN3}, however, does not secure the fulfilment of the self-resonant constraint. This constraint is met by utilizing the linear combination of modes (six modes are degenerated at $\lambda_2 = \lambda_2^\ast$)
\begin{equation}
\V{J} \left( \V{r} \right) = \sum\limits_{q} \alpha_q \V{u}_q \left( \V{r}\right) + \alpha \sum\limits_{r} \alpha_r \V{u}_r \left( \V{r}\right) = \V{J}_\T{e} + \alpha \V{J}_\T{m},
\label{eq:deltaProblem4}
\end{equation}
where~\cite{2017_Losenicky_Comment_Pfeiffer}
\begin{equation}
\alpha = \sqrt{-\dfrac{\langle \V{J}_\T{e} , \OP{X}_0 (\V{J}_\T{e}) \rangle}{\langle \V{J}_\T{m} , \OP{X}_0 (\V{J}_\T{m}) \rangle}} \T{e}^{\J \varphi},
\label{eq:deltaProblem5}
\end{equation}
\cf{}, \eqref{eq:IoptSolution}.
Both dominant TM and TE spherical waves are three-times geometrically degenerated. Therefore, the $\alpha_q$ and $\alpha_r$ parameters are free to choose. 

\subsection{Manipulation With Isolated Eigentraces (Non-Foster Matching Elements)}
\label{subsec:modeManipul}

Recapitulate first that, when symmetries are present, the eigentrace crossings appear and it might be required to combine two (or more) eigenvectors. Dealing with higher-dimensional point groups, more possibilities exist thanks to the geometry degeneracies. The question of whether it is possible to manipulate eigenvalue traces so that more than two traces cross each other and form an additional degree of freedom to constitute an optimal solution is answered in this example. This question goes back to the very first example of Q-factor optimization from Section~\ref{subsec:PTexample} and Section~\ref{subsec:introproblem}.

\begin{figure}
\centering
\includegraphics[]{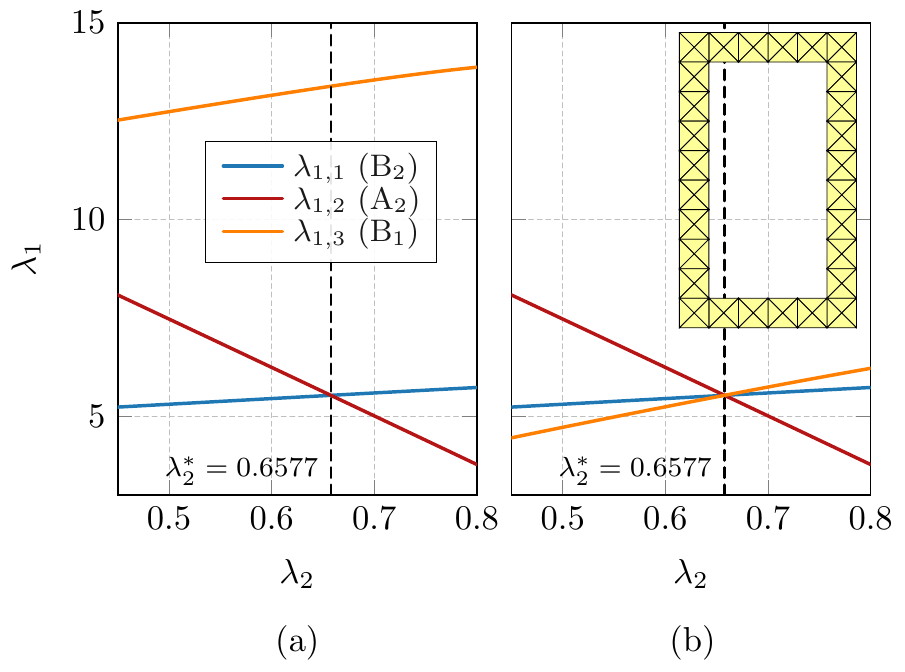}
\caption{Solution to the eigenvalue problem~\eqref{eq:Qsolution2} for a rectangular rim of electrical size $ka = 1$, see the inset. (a) The solution to the original problem~\eqref{eq:Qsolution2}. (b) The solution to the modified problem~\eqref{eq:QdualTuning} with reactance matrix~$\widetilde{\M{X}}_0$ defined so as to manipulate~$\T{B}_1$ irrep only. The irreducible representations of the modes are specified in the legend.}
\label{fig:fig9}
\end{figure}

Let us consider a hypothetical scenario of a passive structure tuned by frequency independent reactances (causing no increase in stored energy). The structure is a PEC rectangular rim of dimensions $\ell \times \ell/2$, see the inset of Fig.~\ref{fig:fig9}b, and of electrical size $ka=1$. Without tuning reactances, the solution to the minimal Q-factor is approached via~\eqref{eq:Qsolution2}, the eigenvalues of which are shown in Fig.~\ref{fig:fig9}a. Let us select the first mode of~$\T{B}_1$ irrep, depicted by the orange curve in Fig.~\ref{fig:fig9}a, and let us find reactance tuning such that this orange trace crosses the other two as depicted in~Fig.~\ref{fig:fig9}b. Notice that the traces~$\lambda_{1,1}(\lambda_2 ),\lambda_{1,2}(\lambda_2 )$ were untouched. In order to manipulate only with traces of irrep~$\T{B}_1$, the reactance matrix of the rim is modified (by the inclusion of frequency independent tuning reactances) as
\begin{equation}
\widetilde{\M{X}}_0 = \Xmvac +  \Gm_{\T{B}_1} \widehat{\M{X}}_\T{L}  \Gm^\T{T}_{\T{B}_1},
\label{eq:X0Tilde}
\end{equation}
where
\begin{equation}
\label{eq:XL}
\widehat{\M{X}}_\T{L} = \mqty[X_{\T{L},1} & \cdots & 0\\
 \vdots & \ddots & \vdots \\ 0 & \cdots & X_{\T{L},N}]
\end{equation}
is a matrix of tuning coefficients. Notice that matrices~$\Gm_{\T{B}_1}$ of the symmetry adapted basis belonging to irrep~$\T{B}_1$ were used in the opposite direction than in relation~\eqref{eq:irrepsAssembly}. This composition guarantees that, irrespective of the matrix~$\widehat{\M{X}}_\T{L}$, only properties of~$\widetilde{\M{X}}_0$ attached to irrep~$\T{B}_1$ will be modified.

\begin{figure}
\centering
\includegraphics[]{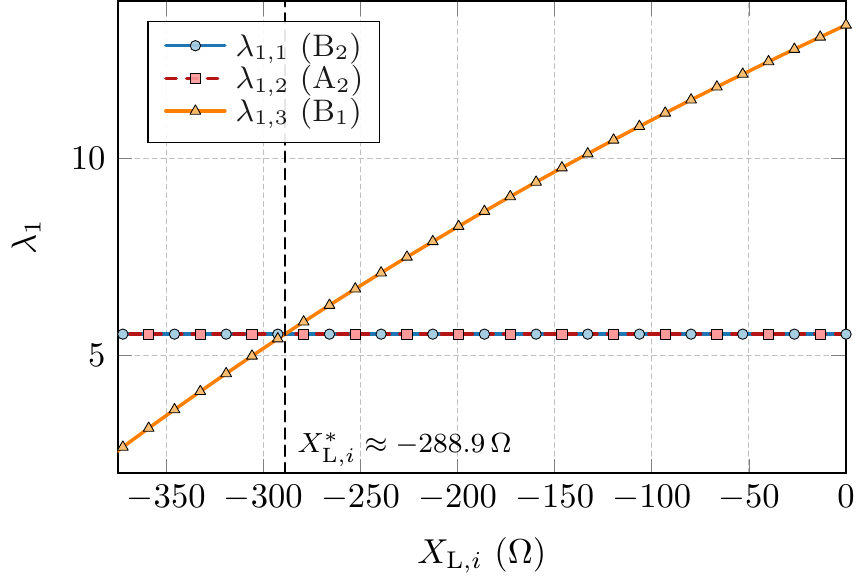}
\caption{Dependence of the dominant eigenvalues~$\lambda_{1,1}$, $\lambda_{1,2}$, and $\lambda_{1,3}$, from irreps $\T{B}_2$, $\T{A}_2$, and $\T{B}_1$, respectively, \cf{}, Fig.~\ref{fig:fig9}, on the tuning parameter~$X_{\T{L},i}$. The matrix~$\widehat{\M{X}}_{\T{L}}$ from~\eqref{eq:XL} is full of zeros, except of the $16$-th position, $i=16$, on the diagonal.}
\label{fig:fig10}
\end{figure}

The use of  reactance matrix~\eqref{eq:X0Tilde} in~\eqref{eq:Qsolution2} instead of matrix~$\M{X}_0$ generates the eigenvalue problem
\begin{equation}
\dfrac{1}{2}\left( \Wm - \lambda_2 \widetilde{\M{X}}_0 \right) \Iv_m = \lambda_{1,m} \Rm \Iv_m,
\label{eq:QdualTuning}
\end{equation}
the results of which are shown in Fig.~\ref{fig:fig10} for the optimal Lagrange's multiplier~$\lambda_2 = \lambda_2^\ast \approx 0.6577$ and for a single non-zero parameter~$X_{\T{L},i} = X_\T{L}$. Notice that particular index~$i$ of a selected tuning parameter is free to choose and is a function of basis functions ordering. 

Orthogonal properties of symmetry adapted bases belonging to different irreps~\eqref{eq:irrepsAssembly} can further be employed to simplify and speed up the evaluation of~Fig.~\ref{fig:fig10}. It has already been mentioned that~\eqref{eq:X0Tilde} cannot change eigentraces belonging to irreps different than~$\T{B}_1$. The blue and red traces in~Fig.~\ref{fig:fig10} are therefore independent of the tuning parameter and there is no need to recalculate them (they attain the same value as in~Fig.~\ref{fig:fig9}). To that point, relation~\eqref{eq:QdualTuning} is left multiplied by~$\Gm^\T{T}_{\T{B}_1}$ and~$\Iv_m = \Gm_{\T{B}_1} \widehat{\Iv}_m$ is substituted which leads to
\begin{equation}
\dfrac{1}{2}\left( \widehat{\Wm}_{\T{B}_1} - \lambda_2^\ast \widehat{\Xm}_{0,{\T{B}_1}} - \lambda_2^\ast \widehat{\M{X}}_{\T{L}} \right) \widehat{\Iv}_{\T{B}_1} = \lambda_{1,m} \widehat{\Rm}_{0,{\T{B}_1}} \widehat{\Iv}_{\T{B}_1}.
\label{eq:QsolutionIrrep1}
\end{equation}
Eigenvalue problem~\eqref{eq:QsolutionIrrep1} generates only those eigensolutions that belong to irrep~$\T{B}_1$ (orange eigentrace in Fig.~\ref{fig:fig10}).

\begin{figure}
\centering
\includegraphics[]{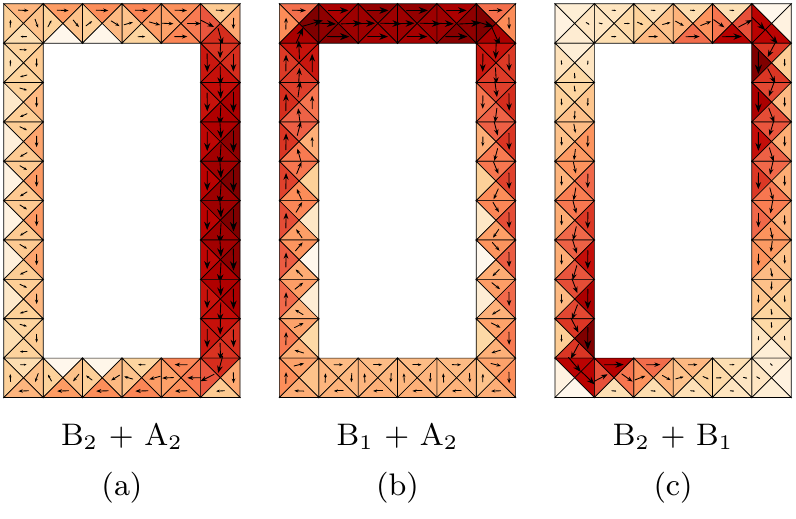}
\caption{Three combinations~\eqref{eq:Iopt} of modes generated by~\eqref{eq:QdualTuning}. Due to the additionally introduced degeneracy in Fig.~\ref{fig:fig9}, not one, but three solutions are possible. The two new solutions are depicted in subfigures (b) and (c).}
\label{fig:fig11}
\end{figure}

It is seen in Fig.~\ref{fig:fig9}b that three eigentraces are crossing each other at~$\lambda_2^\ast$. That means that at least two solutions compliant with~\eqref{eq:Iopt} are possible. These solutions are depicted in Fig.~\ref{fig:fig11} in terms of the resulting surface currents. Case (a) is the classical solution known for~$\T{C}_{2\T{v}}$ combining~$\T{B}_2$ and~$\T{A}_2$ irreps, \cf{}, Fig.~\ref{fig:fig5}c. Case (b) in Fig.~\ref{fig:fig11} is similar to case (a) in shape, but the maximum current density appears on the shorter side. This is possible thanks to added reactive matching elements which effectively elongate the side. Seen in this context, cases (a) and (b) are close to two geometrically degenerated solutions, normally appearing on the square rim ($\T{C}_{4\T{v}}$ point group). Finally, combining solely~$\T{B}$ irreps results in case (c).

In order to reduce the Q-factor in the third irrep, an existence of frequency independent reactance with no energy accumulation was assumed, which is unphysical. A physically more acceptable possibility would be to manipulate the first two modes (from irrep~$\T{A}_2,\T{B}_2$) so they become equal to the third mode. This will rise the Q-factor value, but the gained benefit may be the equality of three eigenvalues (more degrees of freedom). Another possibility is a selective manipulation with a specific sub-set of characteristic modes. The same attempt was already undertaken with geometry manipulations, preserving the symmetries~\cite{YangAdams_SystematicShapeOptimizationOfSymmetricMIMOAntennasUsingCM}, with selective excitation~\cite{SuEtAl_RadiatonEnergyAndMutualCouplingForMIMOAntennaCMs}, or with reactive tuning~\cite{JaafarCollardeySharaiha_OptimizedmanipulationOfNetworkCMsForWidebandSmallAntennaMatching}). With the technique introduced above, one characteristic mode from each irrep can be modified so they all have the same eigenvalue at arbitrary~$ka$. This is possible with simple reactive matching and the procedure above offers a simple recipe of how to do it.

\section{Discussion}
\label{sec:disc}

The determination of fundamental bounds in the presence of symmetries raised several interesting points to be discussed in this section.

\subsection{Robust Algorithm To Eliminate Erroneous Duality Gaps}
\label{subsec:algo}

\begin{figure}
\centering
\includegraphics[width=\columnwidth]{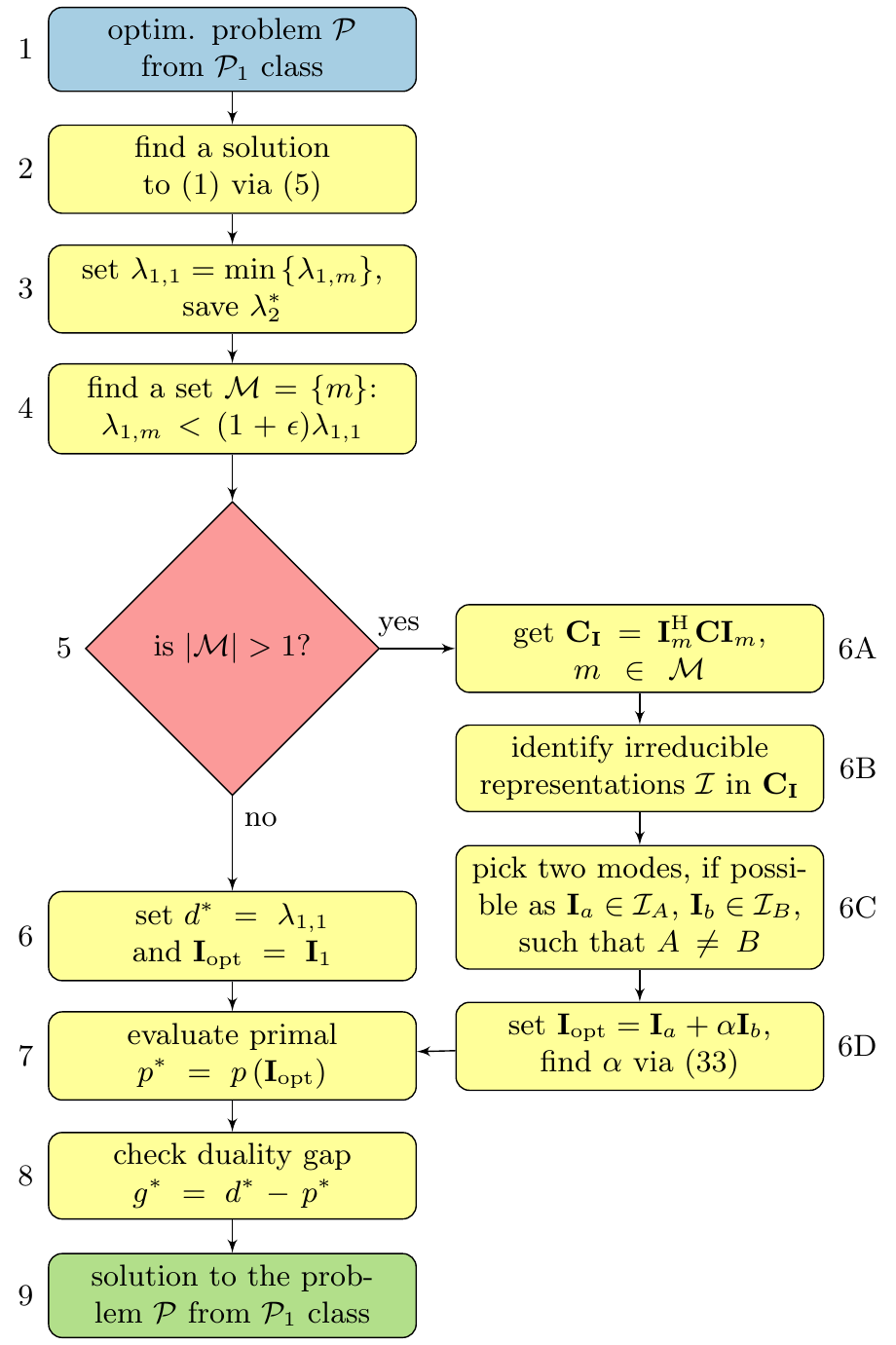}
\caption{Flowchart of a general algorithm dealing with degenerated eigenvalues. It is valid for an arbitrary optimization problem of type~$\OP{P}_1$ (only quadratic constraints) and can deal with the imperfections of the mesh grid, as described in Section~\ref{subsec:mesh}, \cf{} Fig.~\ref{fig:fig4}. In case of geometrical degeneracies within one irrep, see Section~\ref{subsec:geom}, one particular solution is found.}
\label{fig:fig20}
\end{figure}

The procedure capable of dealing with all possible scenarios related to the presence of symmetries is depicted in Fig.~\ref{fig:fig20}. Its robustness was tested against various examples, involving both crossing (mesh grid preserving the symmetries) and near crossing avoidance (slightly unsymmetrical mesh grids), and including shapes from all point groups depicted in Table~\ref{tab:degenerations}.

The workflow is as follows. It is assumed that the problems belonging to class~$\OP{P}_1$ are solved with a dedicated solver (the steps~1 and~2 in Fig.~\ref{fig:fig20}). While the optimal multiplicator~$\lambda_2^\ast$ is found (step 3), identify multiplicity~$|\OP{M}|$ of eigenvalue~$\lambda_1$ (step~4), consider that they may vary up to relative error~$\epsilon$ thanks to numerical errors and mesh imperfections. According to $|\OP{M}|$ decide whether the eigenmodes have to be combined (step~5). Notice that the decision shall not be based on constraint fulfilment as a true duality gap might exist. When degeneracies do not appear, follow standard procedure (steps~6 and~7), \ie{}, determine the value of primal problem~$p^\ast$ (or verify that constraints are fulfilled) and check duality gap~$g^\ast$ (step~8). When the necessity of mode combination is detected, a special routine replacing step~6 is called for (steps~6A--6D). First evaluate projections for a matrix generating one of the constraints (step~6A, matrix~$\M{C}_\M{I}$). Identify block-diagonal matrices within~$\M{C}_\M{I}$ and assign them with different irreps (step~6B). Pick one mode from two different irreps, if not possible, pick two modes arbitrarily (step~6C) and find a value of parameter~$\alpha$ (step~6D).

\subsection{Distinction Between~$\OP{P}_1$-type and~$\OP{P}_2$-type Problems}
\label{subsec:problemDiff}

We have seen that the presence of symmetries has serious consequences for the correct evaluation of problems from class~$\OP{P}_1$. Conversely, problems from class~$\OP{P}_2$ remain untouched. The reason is the presence of a linear term in the constraints which is typically a consequence of a prescribed or, in other words, uncontrollable field quantity. A good example is a prescription for complex power balance, heavily utilized in~\cite{2019_Gustafsson_Arxiv},
\begin{equation}
\Iv^\herm \Zm \Iv = \Iv^\herm \Vv,
\label{eq:complexPower}
\end{equation}
where~$\Vv$ is a vector of excitation coefficients of the incident electric field intensity. Analogous to~\eqref{eq:complexPower} all the linear terms with a current as the unknown couple the optimized quantity to the (external) field. This type of constraint makes the bounds sharper since it connects the optimized quantities and their excitation together.

\subsection{Uniqueness of the Optimal Solution}
\label{subsec:optimSol}

The explicit solution to problems~$\OP{P}_1$ and~$\OP{P}_2$ enlighten the uniqueness of the solutions. In order to simplify the discussion, let us assume that matrices~$\M{A}$ and~$\M{C}$ in~\eqref{eq:problem1} and~\eqref{eq:problem2} have full rank, all the matrices are fixed, and the optimized quantity is properly representable in a basis~\eqref{eq:RWG}, \ie{}, the basis is chosen so that it respects the nature of the optimized problem.

The solution to problem~$\OP{P}_2$ is unique. The solution to non-symmetric problem~$\OP{P}_1$ is non-unique only with respect to the phase of the optimal current. For problem~$\OP{P}_1$ with algebraic multiplicities, as shown in Sec.~\ref{subsec:introproblem}, there is only one value of mixing coefficient~$|\alpha|$, and only the phase of the mixed current may be chosen arbitrarily. Finally, when geometrical multiplicities occur, as shown in Sec.~\ref{subsec:geom}, the optimal current further contains an arbitrary linear combination of geometrically degenerated eigenmodes, see Table~\ref{tab:degenerations}. 

The uniqueness of the optimal currents~$\Iv_\T{opt}$ generating fundamental bounds implies that these bounds are not feasible except for the rare case in which the region used for the optimization,~$\srcRegion$, is already an optimal solution to the shape synthesis problem. Moreover, the excitation used has to be in accordance with the optimal current found, \ie{},
\begin{equation}
\Vv = \Zm \Iv_\T{opt}.
\label{eq:discUniq1}
\end{equation}
If this is not the case, and once the initial shape~$\srcRegion$ has to be perturbed to meet the condition~\eqref{eq:discUniq1}, see, \eg{}, \cite{Capeketal_ShapeSynthesisBasedOnTopologySensitivity}, the removals of the degrees of freedom immediately cause the deterioration of the fundamental bounds, which consequently indicates that the original bound was not feasible.

\begin{table}
\centering
\caption{Maximum number of degenerated eigenvalues depending on the point group of an object (no accidental crossing assumed). Symbols~$A$ and~$B$ represent the dimensions of two irreps of the highest dimensions within the point group. Only the dominant modes are considered.}
\begingroup
\setlength{\tabcolsep}{3pt} 
\renewcommand{\arraystretch}{1} 
\begin{tabular}{cccccc}
 & \includegraphics[width=1.2cm]{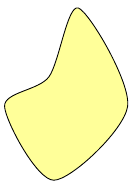} & \includegraphics[width=1.1cm]{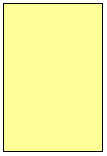} & \includegraphics[width=1.1cm]{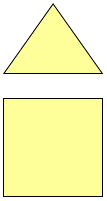} & \includegraphics[width=1.3cm]{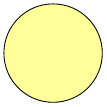} & \includegraphics[width=1.3cm]{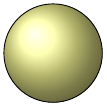} \\ \toprule
point~group & $\T{C}_1$ & $\T{C}_\T{s}$, $\T{C}_\T{2v}$ & $\T{C}_\T{4v}$, $\T{C}_3$ & $\T{O}(2)$ & $\T{O}(3)$ \\
max degen.~$\lambda_1$ & $1+0$ & $1+1$ & $2+1$ & $2+1$ & $3+3$ \\ \bottomrule
\end{tabular}
\endgroup
\label{tab:degenerations}
\end{table}

\subsection{More Than Two Constraints}
\label{subsec:manyConstr}


The existence of a duality gap is not a function of symmetries. Furthermore, the number of constraints and the number of degeneracies are not related in any way. Irrespective of the number of constraints, and considering that there is no duality gap~$g^\ast$, the \Quot{erroneous} duality gap introduced by the presence of symmetries is always eliminated by the proper choice of just one constant, $\alpha$. This statement is explained as follows.

Let us consider a problem from class~$\OP{P}_1$ with multiple constraints leading to the formula for the stationary points~$\Iv$ in a form
\begin{equation}
\M{A} \Iv - \V{\lambda}_2 \M{C} \Iv - \V{\lambda}_3 \M{D} \Iv + \dots =  \V{\lambda}_1 \M{B} \Iv.
\label{eq:discConstr1}
\end{equation}
The dual problem is solved by determining the set of optimal multipliers~$\left\{\lambda_1^\ast, \lambda_2^\ast, \lambda_3^\ast, \dots \right\}$, see step~3 in Fig.~\ref{fig:fig20}. When degeneracies are detected (step~5 in Fig.~\ref{fig:fig20}), a combination of modes from different irreps has to be used (steps~6A--6D in Fig.~\ref{fig:fig20}), introducing an additional degree of freedom, parameter~$\alpha$. Notice that the values of the multipliers, $\left\{\lambda_1^\ast, \lambda_2^\ast, \lambda_3^\ast, \dots \right\}$, are not changed by combining degenerated modal currents. Since we know from the beginning that the problem has no duality gap, \ie{}, the solution to the primal problem is equal to the solution of the dual problem, we know that all the constraints
\begin{equation}
\begin{split}
\Iv_1^\herm \M{B} \Iv_1 + |\alpha|^2 \Iv_2^\herm \M{B} \Iv_2 &= 0 \\
\Iv_1^\herm \M{C} \Iv_1 + |\alpha|^2 \Iv_2^\herm \M{C} \Iv_2 &= 1 \\
\Iv_1^\herm \M{D} \Iv_1 + |\alpha|^2 \Iv_2^\herm \M{D} \Iv_2 &= d \\
\vdots \qquad \qquad &= \, \vdots
\label{eq:discConstr5}
\end{split}
\end{equation}
can be fulfilled by properly setting just one parameter~$\alpha$. This means that we can pick only one of the constraints~\eqref{eq:discConstr5} and determine the proper value of the parameter~$\alpha$.

\section{Conclusion}
\label{sec:concl}

It has been shown that one entire class of optimization problems generating fundamental bounds in electromagnetism is encumbered with potential issues induced by symmetries. When no linear constraints are present, care must be taken with the investigation of the primal solution. This applies to structures with an imperfect discretization mesh grid as well, where the elimination of an erroneous duality gap might be even more problematic since the separation into irreducible representations is not possible. 

A heuristic, yet general and point group theory-based technique to remove the erroneous duality gap has been presented and tested for various examples of varying complexity. The formula was tested for all canonical bodies (rectangular and square plates, a triangular shape, a spherical shell, etc.).

This work helps to understand the role of symmetries in establishing source quantity-based bounds. The challenges related to the presence of symmetries, when properly treated, introduce additional degrees of freedom. All conclusions apply not only to optimal, yet abstract and usually non-realizable, currents but also to optimal port mode excitation and other feasible representations of integro-differential operators.

\appendices

\section{Character Tables for Rectangular and Square Plates ($\T{C}_\T{2v}$ and $\T{C}_\T{4v}$)}
\label{sec:charTables}

This appendix lists character tables of point groups used in the paper. Each table also contains symmetry operations available for the group (those are grouped into conjugacy classes and enumerated in the first row of the table) and the corresponding irreps (enumerated in the first column of the table). The table entries consist of group characters (numbers of the table) and denote the traces of the matrix representations for a corresponding class and irrep. The number of irreps corresponds to the number of classes~\cite{McWeeny_GroupTheory}, all rows and columns of the character table are orthogonal. The symmetry operations used in Tables~\ref{tab:charTableC1}--\ref{tab:charTableC4v} are: $\T{E}$ -- the identical operation, $\sigma_\T{t}$ -- a reflection ($\T{t}$ is a placeholder for a specific type of reflection), $\T{C}_n(u)$ -- a rotation by $2\pi/n$ around $u$ axis, $\sigma_{uv}$ a reflection through plane~$uv$.

The character corresponding to identity operation~$\T{E}$ indicates the dimension of the irrep and the geometric multiplicity of the eigenvalues within that irrep. For example, current solutions falling into the~$\T{E}$ irrep of~$\T{C}_{4\T{v}}$ group are twice degenerated, see Table~\ref{tab:charTableC4v}. This applies to the solutions of the problem~\eqref{eq:Gsolution2} in Section~\ref{subsec:geom}, see Fig.~\ref{fig:fig06}.

\begin{table}[]
\centering
\caption{Character table for point group~$\T{C}_1$, a non-symmetric object belonging to.}
\begin{tabular}{ccc}
$\T{C}_1$ & $\T{E}$ \\ \toprule
$\textsc{A}$ & $+1$ \\ \bottomrule
\end{tabular}
\label{tab:charTableC1}
\end{table}

\begin{table}[]
\centering
\caption{Character table for point group~$\T{C}_\T{s}$, a non-symmetric object over ground plane belonging to.}
\begin{tabular}{ccc}
$\T{C}_\T{s}$ & $\T{E}$ & $\T{\sigma}_\T{h}$ \\ \toprule
$\textsc{A}'$ & $+1$ & $+1$ \\
$\textsc{A}''$ & $+1$ & $-1$ \\ \bottomrule
\end{tabular}
\label{tab:charTableCs}
\end{table}

Non-symmetric objects belong to point group~$\T{C}_1$, see Table~\ref{tab:charTableC1}. Objects with one reflection plane, often classified as having odd and even solutions, belong to point group~$\T{C}_\T{s}$, see Table~\ref{tab:charTableCs}. The remaining two groups mentioned here are~$\T{C}_{2\T{v}}$ (\eg{}, rectangular plate) in Table~\ref{tab:charTableC2v} and~$\T{C}_{4\T{v}}$ (\eg{}, square plate) in Table~\ref{tab:charTableC4v}.

\begin{table}[]
\centering
\caption{Character table for point group~$\T{C}_\T{2v}$, a rectangular plate belongs to.}
\begin{tabular}{ccccc}
$\T{C}_\T{2v}$ & $\T{E}$ & $\T{C}_\T{2}(z)$ & $\T{\sigma}_{xz}$ & $\T{\sigma}_{yz}$ \\ \toprule
$\textsc{A}_1$ & $+1$ & $+1$ & $+1$ & $+1$ \\
$\textsc{A}_2$ & $+1$ & $+1$ & $-1$ & $-1$ \\
$\textsc{B}_1$ & $+1$ & $-1$ & $+1$ & $-1$ \\
$\textsc{B}_2$ & $+1$ & $-1$ & $-1$ & $+1$ \\ \bottomrule
\end{tabular}
\label{tab:charTableC2v}
\end{table}

\begin{table}[]
\centering
\caption{Character table for point group~$\T{C}_\T{4v}$, a square plate belongs to.}
\begin{tabular}{cccccc}
$\T{C}_\T{4v}$ & $\T{E}$ & $2 \T{C}_\T{4}(z)$ & $\T{C}_\T{2}(z)$ & $2\T{\sigma}_{v}$ & $2\T{\sigma}_{d}$ \\ \toprule
$\textsc{A}_1$ & $+1$ & $+1$ & $+1$ & $+1$ & $+1$ \\
$\textsc{A}_2$ & $+1$ & $+1$ & $+1$ & $-1$ & $-1$ \\
$\textsc{B}_1$ & $+1$ & $-1$ & $+1$ & $+1$ & $-1$ \\
$\textsc{B}_2$ & $+1$ & $-1$ & $+1$ & $-1$ & $+1$ \\
$\textsc{E}$ & $+2$ & $0$ & $-2$ & $0$ & $0$ \\ \bottomrule
\end{tabular}
\label{tab:charTableC4v}
\end{table}

\bibliographystyle{ieeetr}
\bibliography{references}

\begin{IEEEbiography}[{\includegraphics[width=1in,height=1.25in,clip,keepaspectratio]{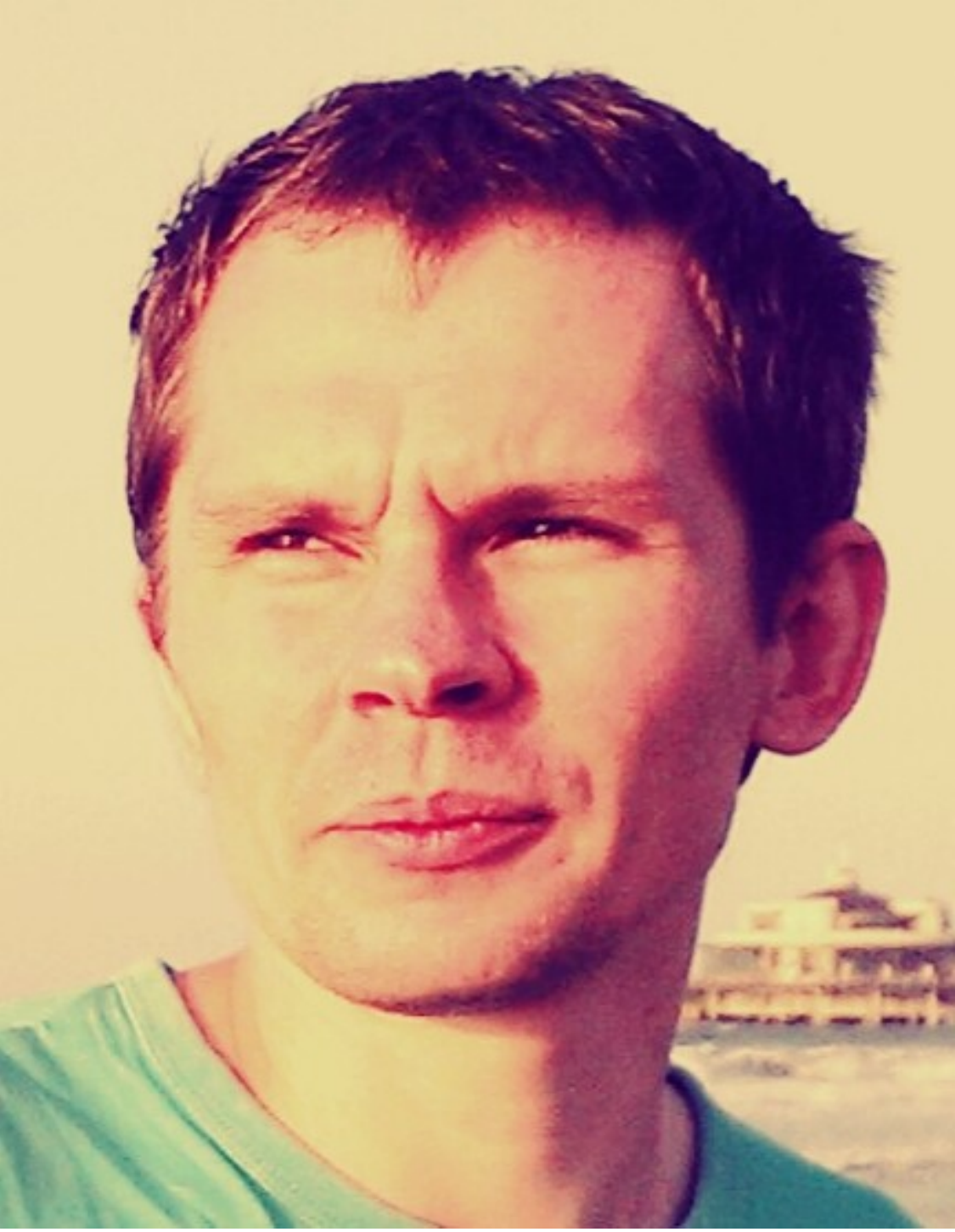}}]{Miloslav Capek}
(M'14, SM'17) received the M.Sc. degree in Electrical Engineering 2009, the Ph.D. degree in 2014, and was appointed Associate Professor in 2017, all from the Czech Technical University in Prague, Czech Republic.
	
He leads the development of the AToM (Antenna Toolbox for Matlab) package. His research interests are in the area of electromagnetic theory, electrically small antennas, numerical techniques, fractal geometry, and optimization. He authored or co-authored over 100~journal and conference papers.

Dr. Capek is member of Radioengineering Society, regional delegate of EurAAP, and Associate Editor of IET Microwaves, Antennas \& Propagation.
\end{IEEEbiography}

\begin{IEEEbiography}[{\includegraphics[width=1in,height=1.25in,clip,keepaspectratio]{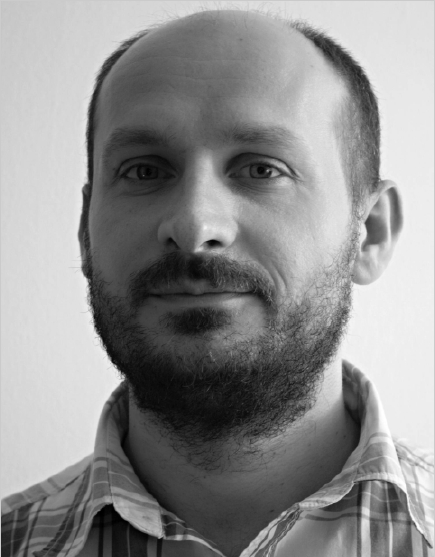}}]{Lukas Jelinek}
received his Ph.D. degree from the Czech Technical University in Prague, Czech Republic, in 2006. In 2015 he was appointed Associate Professor at the Department of Electromagnetic Field at the same university.

His research interests include wave propagation in complex media, general field theory, numerical techniques and optimization.
\end{IEEEbiography}

\begin{IEEEbiography}[{\includegraphics[width=1in,height=1.25in,clip,keepaspectratio]{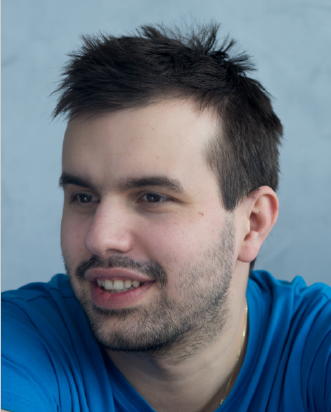}}]{Michal Masek}
received the M.Sc. degree in Electrical Engineering from Czech Technical University in Prague, Czech Republic, in 2015, where he is currently pursuing the Ph.D. degree in the area of modal tracking and characteristic modes. He is a member of the team developing the AToM (Antenna Toolbox for Matlab).
\end{IEEEbiography}

\end{document}